\documentclass[longauth]{aa}  

\usepackage{graphicx}
\usepackage{txfonts}
\usepackage{siunitx}						
\usepackage{physics}
\usepackage{hyperref}
\usepackage{multicol}
\usepackage{lipsum}
\usepackage{ulem}
\usepackage{graphicx}
\usepackage{subcaption}
\usepackage{mwe}
\usepackage[utf8]{inputenc}

\usepackage{xcolor}

\hypersetup{
	colorlinks			= true			, 
	citecolor			= blue!80!black			, 
	linkcolor			= blue			, 
	urlcolor			=blue!80!black		, 
	breaklinks			=true           ,
}

\usepackage{natbib}
\bibpunct{(}{)}{;}{a}{}{,}             
\begin{document}

\title{Improving constraints on the extended mass distribution in the Galactic Center with stellar orbits}
 	
\author{
    The GRAVITY Collaboration\fnmsep\thanks{
    GRAVITY is developed in collaboration by MPE, LESIA of Paris Observatory / CNRS / Sorbonne Université / Univ. Paris Diderot, and IPAG of Université Grenoble Alpes / CNRS, MPIA, Univ. of Cologne, CENTRA - Centro de Astrofisica e Gravitação, and ESO. } 
    : 
    K.~Abd El Dayem             \inst{2}            \and
    R.~Abuter                   \inst{4}            \and
    N.~Aimar                    \inst{2,7}            \and
    P.~Amaro Seoane             \inst{14,1,20,15}   \and
    A.~Amorim                   \inst{8,7}          \and
    J. ~Beck                    \inst{1}            \and
    J.P.~Berger                 \inst{3,4}          \and
    H.~Bonnet                   \inst{4}            \and
    G.~Bourdarot                \inst{1}            \and
    W.~Brandner                 \inst{5}            \and
    V.~Cardoso                  \inst{7,17}         \and
    R.~Capuzzo Dolcetta         \inst{21}           \and
    Y.~Clénet                   \inst{2}            \and
    R.~Davies                   \inst{1}            \and
    P.T.~de~Zeeuw               \inst{}           \and
    A.~Drescher                 \inst{1}            \and
    A.~Eckart                   \inst{6,13}         \and
    F.~Eisenhauer               \inst{1,19}         \and
    H.~Feuchtgruber             \inst{1}            \and
    G.~Finger                   \inst{1}            \and
    N.M.~Förster~Schreiber      \inst{1}            \and
    A.~Foschi                   \inst{2}         \and
    F.~Gao                      \inst{13}         \and
    P.~Garcia                   \inst{10,7}         \and
    E.~Gendron                  \inst{2}            \and
    R.~Genzel                   \inst{1,11}         \and
    S.~Gillessen                \inst{1}            \and
    M.~Hartl                    \inst{1}            \and
    X.~Haubois                  \inst{9}            \and
    F.~Haussmann                \inst{1}            \and
    G.~Heißel                   \inst{16,2}         \and
    T.~Henning                  \inst{5}            \and
    S.~Hippler                  \inst{5}            \and
    M.~Horrobin                 \inst{6}            \and
    L.~Jochum                   \inst{9}            \and
    L.~Jocou                    \inst{3}            \and
    A.~Kaufer                   \inst{9}            \and
    P.~Kervella                 \inst{2}            \and
    S.~Lacour                   \inst{2,4}            \and
    V.~Lapeyrère                \inst{2}            \and
    J.-B.~Le~Bouquin            \inst{3}            \and
    P.~Léna                     \inst{2}            \and
    D.~Lutz                     \inst{1}            \and
    F.~Mang                     \inst{1}            \and
    N.~More                     \inst{1}            \and
    T.~Ott                      \inst{1}            \and
    T.~Paumard                  \inst{2}            \and
    K.~Perraut                  \inst{3}            \and
    G.~Perrin                   \inst{2}            \and
    O.~Pfuhl                    \inst{4,1}          \and
    S.~Rabien                   \inst{1}            \and
    D.~C.~Ribeiro               \inst{1}            \and
    M.~Sadun Bordoni            \inst{1}\thanks{Corresponding author: M.~Sadun Bordoni (mbordoni{@}mpe.mpg.de).}  \and
    S.~Scheithauer              \inst{5}            \and
    J.~Shangguan                \inst{1}            \and
    T.~Shimizu                  \inst{1}            \and
    J.~Stadler                  \inst{12,1}         \and
    O.~Straub                   \inst{1,18}         \and
    C.~Straubmeier              \inst{6}            \and
    E.~Sturm                    \inst{1}            \and
    L.J.~Tacconi                \inst{1}            \and
    I.~Urso                     \inst{2}            \and
    F.~Vincent                  \inst{2}            \and
    S.D.~von~Fellenberg           \inst{13,1}         \and
    F.~Widmann                  \inst{1}            \and
    E.~Wieprecht                \inst{1}            \and
    J.~Woillez                  \inst{4}             \and
    F.~Zhang                    \inst{22,23,24}
}

    \institute{
  	    Max Planck Institute for Extraterrestrial Physics, Giessenbachstraße 1, 85748 Garching, Germany \and
  	    LESIA, Observatoire de Paris, Université PSL, CNRS, Sorbonne Université, Université de Paris, 5 place Jules Janssen, 92195 Meudon, France \and 	
  	    Univ. Grenoble Alpes, CNRS, IPAG, 38000 Grenoble, France \and
  	    European Southern Observatory, Karl-Schwarzschild-Straße 2, 85748 Garching, Germany \and
  	    Max Planck Institute for Astronomy, Königstuhl 17, 69117 Heidelberg, Germany \and
  	    1st Institute of Physics, University of Cologne, Zülpicher Straße 77, 50937 Cologne, Germany \and
  	    CENTRA - Centro de Astrofísica e Gravitação, IST, Universidade de Lisboa, 1049-001 Lisboa, Portugal \and
  	    Universidade de Lisboa - Faculdade de Ciências, Campo Grande, 1749-016 Lisboa, Portugal \and
  	    European Southern Observatory, Casilla 19001, Santiago 19, Chile \and
  	    Faculdade de Engenharia, Universidade do Porto, rua Dr. Roberto Frias, 4200-465 Porto, Portugal \and
  	    Departments of Physics \& Astronomy, Le Conte Hall, University of California, Berkeley, CA 94720, USA \and
  	    Max Planck Institute for Astrophysics, Karl-Schwarzschild-Straße 1, 85748 Garching, Germany \and
  	    Max Planck Institute for Radio Astronomy, auf dem Hügel 69, 53121 Bonn, Germany \and
  	    Institute of Multidisciplinary Mathematics, Universitat Politècnica de València, València, Spain \and
  	    Kavli Institute for Astronomy and Astrophysics, Beijing, China \and
  	    Advanced Concepts Team, ESA, TEC-SF, ESTEC, Keplerlaan 1, 2201 AZ Noordwijk, The Netherlands \and
  	    Niels Bohr International Academy, Niels Bohr Institute, Blegdamsvej 17, 2100 Copenhagen, Denmark \and
  	    ORIGINS Excellence Cluster, Boltzmannstraße 2, 85748 Garching, Germany \and
  	    Department of Physics, Technical University of Munich, 85748 Garching, Germany \and
  	    Higgs Centre for Theoretical Physics, Edinburgh, UK \and
            Department of Physics, Sapienza, University of Rome, P.le A. Moro 5, 00185 Rome, Italy \and
            School of Physics and Materials Science, Guangzhou University, Guangzhou 510006, People's Republic of China \and
            Key Laboratory for Astronomical Observation and Technology of Guangzhou, 510006 Guangzhou, People's Republic of China \and
            Astronomy Science and Technology Research Laboratory of Department of Education of Guangdong Province, Guangzhou 510006, People's Republic of China
    }  
    \date{September 18, 2024}
    \abstract{
 Studying the orbital motion of stars around Sagittarius A* in the Galactic Center provides a unique opportunity to probe the gravitational potential near the supermassive black hole at the heart of our Galaxy. Interferometric data obtained with the GRAVITY instrument at the Very Large Telescope Interferometer (VLTI) since 2016 has allowed us to achieve unprecedented precision in tracking the orbits of these stars. GRAVITY data have been key to detecting the in-plane, prograde Schwarzschild precession of the orbit of the star S2, as predicted by General Relativity. By combining astrometric and spectroscopic data from multiple stars, including S2, S29, S38, and S55 - for which we have data around their time of pericenter passage with GRAVITY - we can now strengthen the significance of this detection to an approximately $10 \sigma$ confidence level.
 The prograde precession of S2's orbit provides valuable insights into the potential presence of an extended mass distribution surrounding Sagittarius A*, which could consist of a dynamically relaxed stellar cusp comprised of old stars and stellar remnants, along with a possible dark matter spike. Our analysis, based on two plausible density profiles - a power-law and a Plummer profile - constrains the enclosed mass within the orbit of S2 to be consistent with zero, establishing an upper limit of approximately $1200 \, M_\odot$ with a $1 \sigma$ confidence level. This significantly improves our constraints on the mass distribution in the Galactic Center. Our upper limit is very close to the expected value from numerical simulations for a stellar cusp in the Galactic Center, leaving little room for a significant enhancement of dark matter density near Sagittarius A*.
    }
  

    \titlerunning{
    Improving constraints on the extended mass distribution in the Galactic Center with stellar orbits}
    \authorrunning{GRAVITY Collaboration}
    \maketitle


\section{Introduction}
\label{introduction}

Since 2016, the GRAVITY interferometer at ESO's Very Large Telescope \citep{GRAVITY_2017} has allowed to obtain astrometric data with unprecedented accuracy (reaching in the best cases a $1 \sigma$ uncertainty of $30 \, \mu as$) of the S-stars orbiting around Sagittarius A* (Sgr A*) in the Galactic Center (GC). This has turned them into a powerful tool to investigate the gravitational potential near the supermassive black hole (SMBH) at the center of our Galaxy, reaching  distances from Sgr A* down to about a thousand times its Schwarzschild radius ($R_S$). Furthermore, astrometric and polarimetric observations of flares from Sgr A* with GRAVITY have revealed that the mass inside the flares’ radius of a few $R_S$ is consistent with the black hole mass measured from stellar orbits \citep{GRAVITY_2018b, GRAVITY_2023_flares}. This, together with the radio-VLBI image of Sgr A* \citep{EHT_2022}, have confirmed that Sgr A* is a SMBH beyond any reasonable doubt.\\
For the S2 star, due to its short orbital period of 16 years and its brightness ($m_K \approx 14$), 
astrometric data are available for two complete orbital revolutions around Sgr A*, while spectroscopic data cover one and a half revolutions \citep{Schodel_2002, Ghez_2003, Ghez_2008, Gillessen_2017}.
At pericenter S2 reaches a distance of $\sim 1400 \, R_S$ from the SMBH with a speed of $7700 \, km \, s^{-1} \simeq 0.026 \, c$.
Monitoring the star's motion on sky and radial velocity with GRAVITY and SINFONI around the time of the pericenter passage in 2018, crucial data were obtained in order to detect the first-order effects in the post-Newtonian (PN) expansion of General Relativity (GR) on its orbital motion.
The first one is the gravitational redshift of spectral lines, which was detected together with the transverse Doppler effect, predicted by Special Relativity, with a $\approx 10 \sigma$ significance in \cite{GRAVITY_2018} and a $\approx 5 \sigma$ significance in \cite{Do_2019a}. \cite{GRAVITY_2019} improved the significance of the detection to $\approx 20 \sigma$. 
The other effect is the prograde, in-plane precession of the orbit's pericenter angle, namely the Schwarzschild precession (SP). It corresponds to an advance of $\delta \varphi_{Schw} = \frac{3 \pi R_S}{a(1-e^2)}$ per orbit, which for S2 is equal to $12.1$ arcminutes per orbit in prograde direction. 
In \cite{GRAVITY_2020a} this effect was detected at the $5 \sigma$ level, and improved in \cite{GRAVITY_2022} to $\approx 7 \sigma$ by combining the data of S2 with data of the stars S29, S38, and S55, that could be observed around their time of pericenter passage with GRAVITY and whose pericenter distances are comparable to that of S2. \\
The Lense-Thirring effect, caused by the spin of the central SMBH, appears at 1.5PN order and gives both an additional contribution to the in-plane precession and a precession of the orbital plane \citep{Merritt_2010}. We define $A_{LT} = 4 \pi \chi \left(\frac{R_S}{2a(1-e^2)} \right)^{3/2}$, which for S2 is equal to $0.11$ arcminutes. Consequently, the in-plane precession per orbit becomes $\delta \varphi_{Kerr} = \delta \varphi_{Schw} -2 A_{LT} \, \textrm{cos}(i)$, while the precession per orbit of the orbital plane is given by $\delta \Phi_{Kerr}= A_{LT}$, where $\chi$ is the dimensionless spin of the SMBH (with $0\leq \chi \leq 1$) and $i$ is the angle between the direction of the SMBH spin and that of the stellar orbital angular momentum. The effect is thus at least 50 times smaller than the SP, assuming a SMBH with maximum spin, and is out of reach for current measurements. In order to measure the spin of Sgr A*, we would need to observe a star with a pericenter distance that is at least three times smaller than that of S2, given the astrometric accuracy achievable with GRAVITY \citep{Waisb_2018, Sad_2023}.\\
Any extended mass distribution around Sgr A*, following a spherically symmetric density profile, would add a retrograde precession of the stellar orbits, counteracting the prograde Schwarzschild precession \citep{GRAVITY_2020a, GRAVITY_2022}. 
This mass distribution is expected to be composed mainly of a dynamically relaxed cusp of old stars and stellar remnants.
\cite{Peebles_1972, frank_rees_1976, BW_1976_1} first addressed the problem of the distribution of stars around a central massive BH. \cite{BW_1976_1} found that a single-mass stellar population around a central massive BH reaches a stationary density distribution over the two-body relaxation timescale, which is a power-law $\rho(r) \propto r^{s}$ with slope $s=-1.75$. 
In the GC, the old stellar population can be approximately represented by light stars with masses around $1 \, M_\odot$ and heavier stellar black holes with masses around $10 \, M_\odot$ \citep{Tal_2017}.
For such a population mass segregation occurs, where heavier objects tend to concentrate towards the center due to dynamical interactions with lighter objects. The mass-segregation solution for the steady-state distribution of stars around a massive BH is derived in \cite{Tal_2009}. It has two branches, weak and strong segregation, based on the dominance of the heavier or lighter objects in the scattering interactions. In the weak segregation branch, the heavy objects settle into a power-law distribution with a slope of $-1.75$, while the lighter objects exhibit a shallower profile with slope $-1.5$, as was already heuristically derived in \cite{BW_1977_2}. Conversely, the strong segregation branch results in steeper slopes and a larger difference between the light and heavy masses. The heavy masses settle into a much steeper cusp with $ -2.75 \lesssim s \lesssim -2 $, while the light masses into a cusp with  $ -1.75 \lesssim s \lesssim -1.5$. \cite{Preto_Pau_2010} have provided a clear realization through N-body simulations of the strong mass segregation solution, showing also that the stellar cusp can develop on timescales that are much shorter than the relaxation time, which is shorter than the Hubble time for the GC \citep{Tal_2009, Genzel_2010}. In \cite{Linial_2022} it is argued that weak segregation must exist interior to a certain break radius $r_B$ where the massive population dominates the scattering, while for radii larger than $r_B$ the light objects dominate the scattering and strong segregation occurs.\\
The existence of such a stellar cusp in the GC is also validated by the observational results of \cite{Schodel_2018_1} and \cite{Schodel_2018_2} for the distribution of giant, sub-giant and main-sequence stars within the central few parsecs. They find that the density distribution of the light objects is shallower than $s=-1.5$, being compatible with a power-law with slope between $-1.4$ and $-1.15$.
This is impossible in the steady state, Bahcall \& Wolf framework in order to maintain an equilibrium distribution, but could be explained by a number of factors, such as stellar collisions \citep{rose_2023}, taking into account the complex star formation history of the nuclear star cluster \citep{Baumb_2018}, or by diffusion in angular momentum leading to tidal disruptions, namely diffusion into the loss cone \citep{zhang_2023}.
Red giant stars, instead, do not show a cusp but a distribution which appears to flatten towards the central $\sim 0.3 \, pc$ \citep{Buchholz_2009, Do_2009, Barkto_2010, Schodel_2018_1}, possibly due to the stripping of red giant envelopes due to the interaction with a star-forming disk \citep{Amaro-Seoane_2014}.
\\
In addition to the stellar cusp, an intermediate mass black hole (IMBH) companion of Sgr A* could be present in the GC. It has been shown that an IMBH enclosed within the orbit of S2 can only have a mass $<10^3 \, M_\odot$ \citep{GRAVITY_2023a, Will_2023}. Moreover, it was predicted by \cite{Silk_99} that dark matter particles could be accreted by the SMBH to form a dense spike, increasing the dark matter density in the GC up to ten orders of magnitude with respect to the expected density in case of a Navarro-Frenk-White (NFW) profile. 
In this scenario, the spike could contribute to the extended mass distribution around Sgr A*, while in the absence of such a spike, the contribution of dark matter within the radial range of the S-stars' orbits would be negligible under an NFW profile.
The dark matter spike would also follow a power-law distribution $\rho(r) \propto r^{s}$, with slope $ -2.5 < s < -2.25$ in case of a generalized NFW profile \citep{Silk_99, Shen_2023}. Another possibility that has been investigated is that dark matter could exist in the form of an ultralight scalar field or a massive vector field cloud that clusters around Sgr A* \citep{GRAVITY_2023_c, GRAVITY_2024}, or as a compact fermion ball supported by degeneracy pressure \citep{viollier_1993,arguelles_2019, Becerra_Vergara_2020}. 
\\
In \cite{GRAVITY_2022} the $1 \sigma$ upper limit on any extended mass distributed within the orbit of S2 was found to be $\approx 3000 M_\odot$, assuming a Plummer density profile \citep{Plummer_1911}.
In this paper we use S-stars data, including one more year of GRAVITY observations, in order to improve and extend the analysis conducted in \cite{GRAVITY_2022}. Our goal is to enhance the significance of the Schwarzschild precession detection and to improve the observational constraints on the extended mass distribution around Sgr A*, comparing these results with theoretical models.
In Section \ref{obs} we describe our data set. In Section \ref{Schwarzschild} we give our results on the Schwarzschild precession detection and in Section \ref{extmass} on the extended mass distribution. In Section \ref{conclusions} we summarize our conclusions.

\section{Observations}
\label{obs}

We described in \cite{GRAVITY_2022} that interferometric astrometry with GRAVITY gives many advantages over single-telescope AO imaging. Most importantly, it allows us to reach a much higher angular resolution and astrometric accuracy and to be significantly less affected by confusion noise. In this paper we add one more year of GRAVITY data (up to fall of 2022) for the stars S2, S29, S38 and S55 with respect to \cite{GRAVITY_2022}. 
In addition, since April 2023 the Enhanced Resolution Imager and Spectrograph (ERIS), has started to be operational at the VLT \citep{ERIS_2023}. This has made possible, after the decommissioning of SINFONI \citep{Eisenhauer_2005} in 2019, to resume spectroscopic observations of the S-stars, with a new state-of-the-art AO system and a much higher spectral resolution ($R \sim 10000$, compared to $R \sim 4000$ of SINFONI). In this paper we add to the long-standing SINFONI data set the first radial velocity data points obtained with ERIS, during the commissioning phase in 2022.\\
For the analysis conducted in this paper we used the following data set:
\begin{itemize}
    \item For S2 we used 128 NACO and 82 GRAVITY data points for the astrometry, and 92 SINFONI, 3 Keck, 2 NACO, 4 GNIRS / GEMINI and 3 ERIS radial velocity measurements. These data cover the time span of 1992.2-2022.7.
    \item For S29 we used 66 NACO (2002.3-2019.7) and 29 GRAVITY data points (2019.6-2022.7) for the astrometry. We dropped a significant fraction of the available NACO data, which have been affected by confusion events. We also use 17 SINFONI and 2 GNIRS radial velocity measurements (2007.6-2021.4).
    \item For S38 we used 110 NACO (2004.2-2018.6) and 23 GRAVITY (2021.2-2022.7) data points for the astrometry, and 8 SINFONI, 2 Keck and 1 ERIS data points for the radial velocity (2008.3- 2022.4).
    \item For S55 we used 42 NACO (2004.5-2013.6) and 27 GRAVITY (2021.2-2022.6) astrometric data points, and 2 SINFONI radial velocity data points (2014).
    \item We also used the NACO and GRAVITY astrometric data and the SINFONI radial velocity data for other two stars: S24 and S42. For five other stars, S1, S4, S9, S21 and S13, we only used NACO and SINFONI data.
\end{itemize}

Figure \ref{fig_orb} illustrates the orbits of all these stars. Details on the analysis of GRAVITY data can be found in Appendix A of \cite{GRAVITY_2020a}. The analysis of spectroscopic data is described in \cite{Habibi_2017} and \cite{GRAVITY_2019}.

   \begin{figure}
   \centering
    \includegraphics[width=0.5\textwidth]{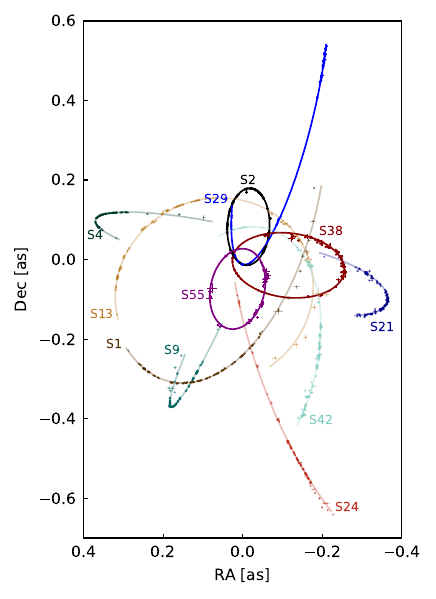}
   \caption{Orbits for the set of 11 S-stars that have been used in this paper. Highlighted are the 4 most relevant stars to constrain the gravitational potential around Sgr A*: S2, S29, S38 and S55.}
              \label{fig_orb}
    \end{figure}


\section{Schwarzschild precession of the orbit of S2}
\label{Schwarzschild}

\subsection{Method}
\label{sec:analysis_schw}
To test whether the orbits of S-stars are well described by a Schwarzschild orbit around Sgr A* we model their acceleration as given by a first-order PN approximation of GR for a massless test particle \citep{Will_1993} and multiply the 1PN terms by a factor $f_{SP}$, such that $f_{SP}=0$ corresponds to a Keplerian closed (non-precessing) orbit and $f_{SP}=1$ to a GR Schwarzschild orbit, with a prograde precession of the orbit's pericenter angle.
The parameter $f_{SP}$ is then used as a fitting parameter, together with the mass of and distance to the central SMBH, $m_\bullet$ and $R_0$, five coordinates ($x_0, y_0, vx_0, vy_0, vz_0$) describing the position on sky and three-dimensional velocity of Sgr A* in the AO imaging/spectroscopic reference frame, and six orbital parameters $(a, e, i, \omega, \Omega, t_0)$ per each star that we use in the orbital fit, describing the initial osculating Kepler orbit.\\
The orbital fitting is done using either a Levenberg-Marquardt $\chi^2$ minimization algorithm or a (Metropolis-Hastings) Markov-chain Monte Carlo (MCMC) analysis, using $200 \, 000$ realizations. For more details, see \cite{GRAVITY_2018, GRAVITY_2019, GRAVITY_2020a}.

    \begin{figure*}
        \centering
        \begin{subfigure}{\textwidth}
            \centering
            \includegraphics[width=0.8\textwidth]{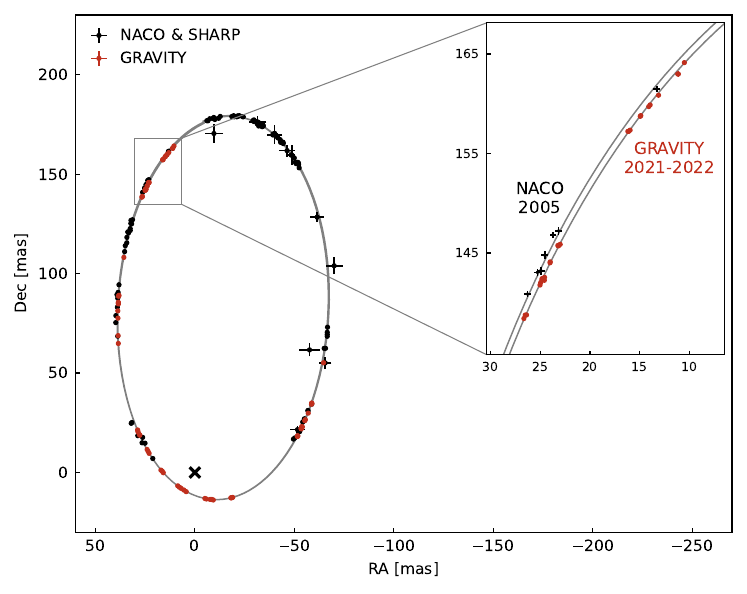}
        \end{subfigure}
        \hfill
        \begin{subfigure}{\textwidth}  
            \centering 
            \includegraphics[width=0.8\textwidth]
            {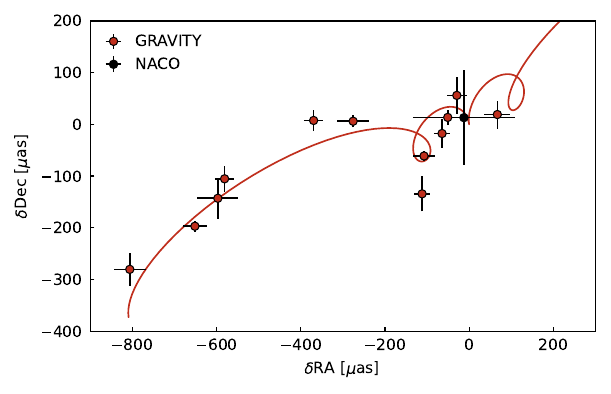}
        \end{subfigure}
        \caption[]
        {Top: Astrometric data of S2 obtained from 1992 to the end of 2022 together with the best-fitting orbit. The SHARP and NACO data are in black, while the GRAVITY data are in red. Sgr A* is located at (0,0), marked by a black cross. A zoom-in shows the effect of the prograde, Schwarzschild precession, comparing the GRAVITY 2021-2022 data with the NACO 2005 data. \\
        Bottom: Residuals in RA and Dec between the best-fit Schwarzschild orbit ($f_{SP} = 1$) and the Newtonian component of the same orbit ($f_{SP} = 0$). In the plot the Schwarzschild orbit predicted by GR corresponds to the loopy red curve. The corresponding residual GRAVITY data with respect to the Newtonian orbit are represented by red circles, the NACO data by a black circle. The GRAVITY data points follow the Schwarzschild orbit predicted by GR.
        GRAVITY data are averages of several epochs, the NACO data have been averaged into one single data point.} 
        \label{fig:s2}
    \end{figure*}

\subsection{Results}
\label{sec:results_schw}
Here we use the complete data set of the stars S2, S29, S38 and S55 (Section \ref{obs}) to obtain the best possible constraint on $f_{SP}$.
As noted in \cite{GRAVITY_2020a, GRAVITY_2022}, the NACO zero points $x_0, y_0, vx_0, vy_0$ are partially degenerate with $f_{SP}$. To mitigate this degeneracy, we use data from seven additional S-stars (see Section \ref{obs} and Figure \ref{fig_orb}), performing a combined Keplerian fit to obtain an estimate on $x_0, y_0, vx_0, vy_0$. 
We then combine the result on $x_0, y_0, vx_0, vy_0$ obtained from the stellar orbits with the constraints from the NACO astrometry of flares and the prior from the construction of the AO reference frame \citep{Plewa_2015}, in order to derive a prior on these parameters (see Appendix \ref{fit_details}).\\
Fitting the data of S2 using this prior for the NACO zero points we obtain $f_{SP}=0.918 \pm 0.128$ (MCMC $f_{SP}=0.911 \pm 0.131$).
The result improves when fitting the data of S2 together with S29, S38 and S55. In fact, these stars passed through pericenter between 2021 and 2023, allowing us to observe them around their time of pericenter passage with GRAVITY \citep{GRAVITY_2022}. Their pericenter distances are comparable to that of S2, ranging from $1200 \, R_S$ to $2800 \, R_S$, providing crucial data to improve our constraints on the gravitational potential around Sgr A*. With this comprehensive dataset, we obtain $f_{SP}=1.135 \pm 0.110$ (MCMC $f_{SP}=1.133 \pm 0.113$). The significance of the Schwarzschild precession detection has thus increased from $\approx 7 \sigma$ of \cite{GRAVITY_2022} to $\approx 10 \sigma$. It is therefore strikingly evident that the orbit of S2 is, with ever stronger significance, described by a Schwarzschild orbit around Sgr A*, exhibiting a prograde, in-plane precession of its pericenter angle. In the upper panel of Figure \ref{fig:s2}, we show the NACO and GRAVITY astrometric data of S2, highlighting the effect of the Schwarzschild precession by comparing the GRAVITY 2021-2022 data with the NACO data from 2005, obtained one orbital period earlier. In the bottom panel we show the averaged S2 residual data obtained with GRAVITY and NACO, after subtracting the Newtonian component of the best-fit Schwarzschild orbit. The data points follow the Schwarzschild orbit predicted by GR, which corresponds to the loopy red line in the plot.\\
For the first time, we are also able to measure the Schwarzschild precession of S2's orbit using only the GRAVITY astrometric data and the radial velocity data, due to a coverage of almost half of S2's orbit with GRAVITY. This allows us to completely exclude the NACO data from the fit and remove the NACO reference frame parameters $x_0, y_0, vx_0, vy_0$, since with GRAVITY we can directly measure the separation vector between Sgr A* and S2. Fitting the orbit of S2 we obtain $f_{SP}=1.016 \pm 0.226$ (MCMC $f_{SP}= 1.024 \pm 0.260$), indicating a measurement of the Schwarzschild precession with a $\approx 4\sigma$ confidence. This leads to a less significant detection than that achieved by including the NACO data in the analysis, despite fitting four fewer parameters, which helps to reduce some degeneracies. In Appendix \ref{sp_prediction} we attempt to predict how much we will be able to improve our constraint on $f_{SP}$ by continuing to monitor the orbit of S2 with GRAVITY and ERIS, carrying out mock observations. We show that GRAVITY data taken in the following years, until the S2 star will reach the next apocenter passage in 2026, will only moderately improve the SP detection. Additionally, we show that the NACO imaging data will still be key in the near future in order to achieve the best possible constraint on $f_{SP}$, at least until S2 will have gone through the next pericenter passage in 2034.

\section{The extended mass distribution around Sgr A*}
\label{extmass}

\subsection{Method}
\label{sec:analysis_extmass}

To test the existence of an extended mass distribution around Sgr A* we fix $f_{SP}=1$ (we assume a 1PN approximation of GR) and we model the extended mass distribution using a spherically symmetric density profile $\rho (r)$. In this case, the motion of stars is not only due to the SMBH gravitational potential but also to the potential generated by this additional mass distribution, which gives an additional term to the stars acceleration and causes a retrograde precession of their orbits.\\
We choose to test two plausible density profiles:
\begin{itemize}

    \item The power-law profile
    \begin{equation}
    \rho(r)=  \rho_0 \left( \frac{r}{r_0} \right)^{s},
    \end{equation}
    where $r_0$ is a length scale for the radial coordinate (we fix it to $r_0= 40$ mpc), $\rho_0$ is the density at $r=r_0$, and $s$ is the slope of the power-law.
    Integrating we obtain the enclosed mass as a function of radius: 
     \begin{equation}
        m(r)= \frac{4 \pi \rho_0}{3+ s} \left( \frac{r^{3+s}}{r_0^{s}} \right).
    \end{equation}

    This is the distribution expected in case of a stellar cusp, with slope $-2.75 \lesssim s \leq -1.5$ depending on the mass of the stellar population and whether weak or strong mass segregation occurs \citep{Tal_2009, Preto_Pau_2010}, and in case of a dark matter spike, with slope $-2.5 < s < -2.25$ \citep{Shen_2023}.
    We investigate this profile varying the slope in the range $-3 < s \leq 0$. We thus explore a range of possible slopes for the density distribution, from a constant density profile $\rho=\rho_0=const$ for $s=0$, to progressively steeper distributions with divergent central density for $s<0$. \\
    
    \item The Plummer profile \citep{Plummer_1911}
\begin{equation}
    \rho(r)= \rho_0 \left[ 1+ \left( \frac{r}{a} \right)^2\right]^{-5/2},
\end{equation}
    where $a$ is the scale radius of the distribution, such that $\rho(r=a) = \frac{\rho_0}{2^{5/2}}$, and $\rho_0$ is the (finite) central density. \\
    Integrating we obtain the enclosed mass profile:
    \begin{equation}
        m(r)=  \frac{4}{3}\pi \rho_0 a^3 \frac{(r/a)^3}{\left[1+ (r/a)^2 \right]^{3/2}}.
    \end{equation}
    The total mass of the distribution is equal to $m_{tot}=\frac{4}{3}\pi \rho_0 a^3$.
    We investigate this profile varying the scale radius in the range $0< a \leq 40$ mpc, going from very compact to progressively broader distributions, exploring thus also the possibility that the density distribution reaches a plateau (a finite value $\rho(r=0)=\rho_0)$ at the center.
\end{itemize}

The procedure we adopt is the following. We parametrize both distributions by defining $\rho_0= f_{pow} m_\bullet$ for the power-law and $\rho_0= \frac{f_{pl} m_\bullet}{\frac{4}{3}\pi a^3}$ for Plummer.
In the first case, we fix the slope $s$ of the power-law to different values in the interval (-3, 0] and fit for the parameter $f_{pow}$, again together with $m_\bullet$, $R_0$, the reference frame parameters $x_0, y_0, vx_0, vy_0, vz_0$ and six orbital parameters $(a, e, i, \omega, \Omega, t_0)$ per each star.
In the second case, we fix the scale radius $a$ of the Plummer profile to different values in the interval (0, 40] mpc and fit for $f_{pl}$ together with the other parameters.  The orbital fitting is done using an MCMC analysis with $200 \, 000$ realizations.
We do not allow the parameters $s$ (for the power-law) and $a$ (for Plummer) to vary freely in the fit alongside $f_{pow}$ and $f_{pl}$ because doing so significantly increases the computational cost of the MCMC analysis, making it challenging to effectively explore the entire parameter space.
It is assumed as a prior that $f_{pow} \geq 0$ and $f_{pl} \geq 0$, ensuring that the density $\rho(r) \geq 0$ for every value of r.
Then we convert the resulting posterior distribution of the parameters $f_{pow}$ and $f_{pl}$ into a distribution on the parameter $M_{encl, S2}= m(r_{peri,S2} < r < r_{apo,S2})$, namely the enclosed mass within the orbit of S2 (see Appendix \ref{fit_details}), where the pericenter distance of S2 is $r_{peri,S2} \sim 0.6$ mpc and the apocenter distance is $r_{apo,S2} \sim 9.4$ mpc.
We focus on the enclosed mass within S2's orbit because it corresponds to the radial range over which the data used in the orbital fits lie. Presenting the results on the total mass of the Plummer distribution or the total mass of the power-law distribution within a radius $r_{cut}>>r_{apo,S2}$ is less relevant, as they depend on the chosen model for the density distribution and cannot be effectively constrained by our data. In addition, the mass enclosed within S2 pericenter $m(r_{peri,S2})$ is degenerate with the SMBH mass $m_\bullet$, and for very steep distributions it is not well constrained with our data (see Appendix \ref{fit_details}).

\subsection{Results}
\label{sec:results_extmass}
\begin{figure*}
        \centering
        \captionsetup[subfigure]{position=top}
        \begin{subfigure}[b]{0.48\textwidth}
            \centering
            \caption[]
            {\textbf{ Power-law profile}}
            \includegraphics[width=\textwidth]{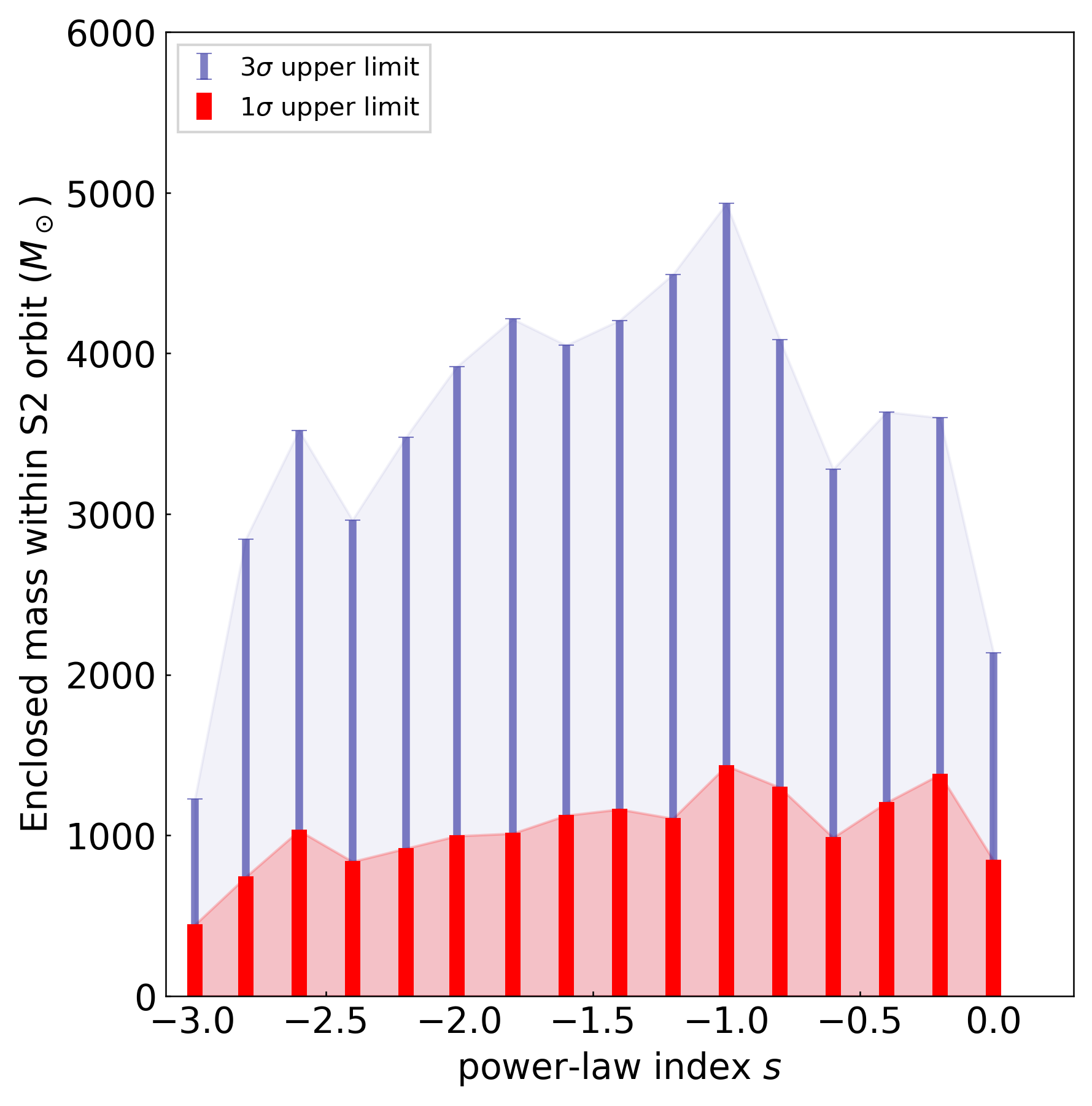}
        \end{subfigure}
        \hfill
        \begin{subfigure}[b]{0.48\textwidth}  
            \centering 
            \caption[]%
            {\textbf{Plummer profile}}
            \includegraphics[width=\textwidth]{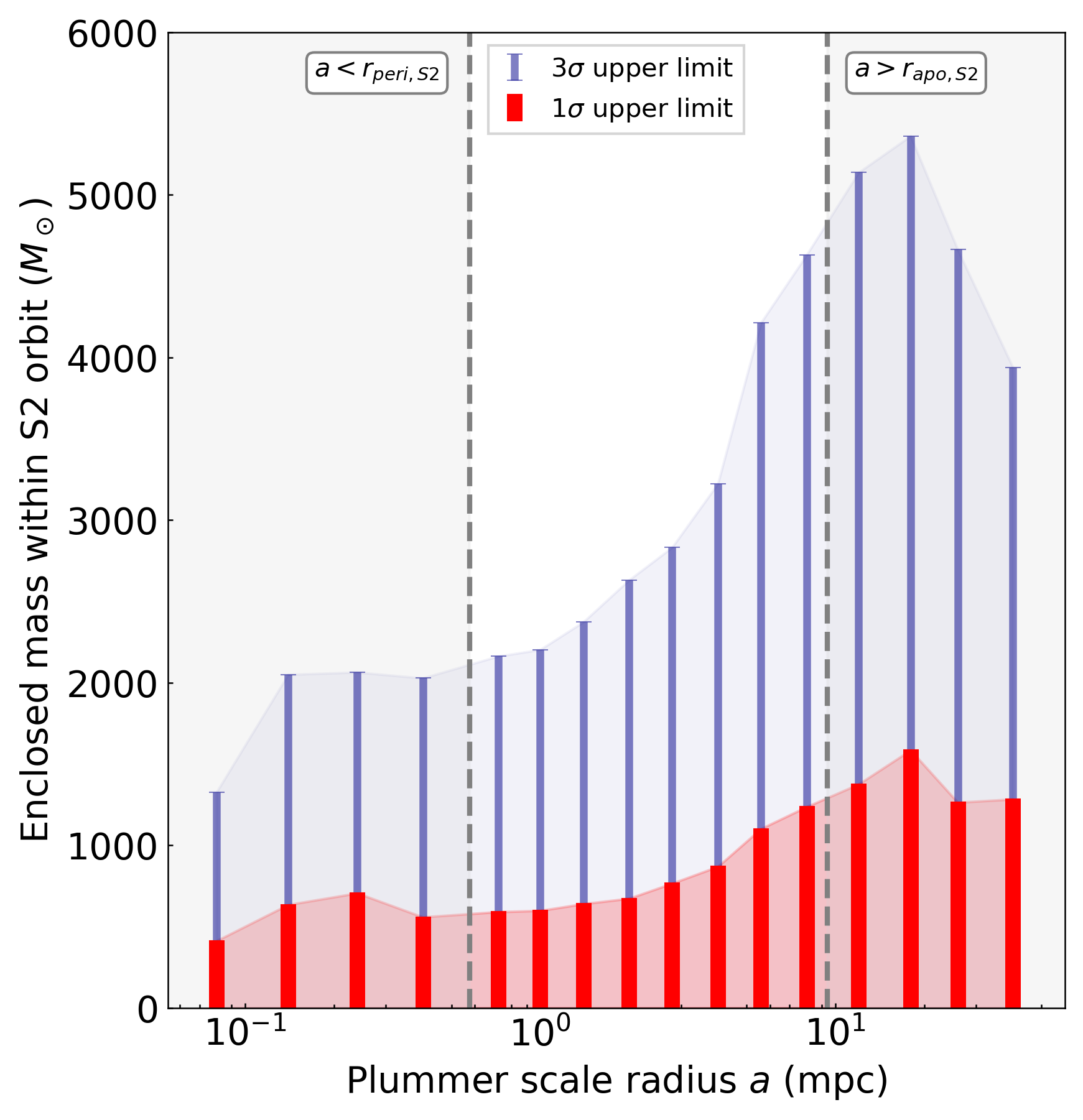}
        \end{subfigure}
        \caption[]
        {Upper limit on the enclosed mass within the orbit of S2 for an extended mass distribution around Sgr A*. We test two plausible density profile for the mass distribution: a power-law profile with varying slope (panel a) and a Plummer profile with varying scale radius (panel b). In red we plot the $1 \sigma$ upper limit and in blue the $3 \sigma$ upper limit, derived from a multi-star fit using data from the stars S2, S29, S38 and S55. Independently of the density profile, the enclosed mass within S2’s orbit is consistently compatible with zero. We set a strong upper limit of approximately $1200 \, M_\odot$ with a $1 \sigma$ confidence level, for reasonable choices of the slope of the power-law ($s<-1.2$) and the scale radius of the Plummer profile ($a \lesssim 8$ mpc).
        } 
        \label{fig:extmass_4s}
    \end{figure*}

The fact that we observe a prograde, in-plane precession of S2's orbit as predicted by GR implies that we can get strong constraints on the potential existence of an extended mass component distributed around Sgr A*. In fact, if such an extended mass component were present, it would induce a retrograde precession of S2’s orbit, counteracting the prograde Schwarzschild precession. \\
To test for the existence of an extended mass distribution around Sgr A*, we follow the procedure described in Section \ref{sec:analysis_extmass}, performing a multi-star fit with the stars S2, S29, S38, and S55. For each star, we utilize the complete data set available (Section \ref{obs}), using both NACO and GRAVITY astrometric data. We impose a prior on the NACO zero points $x_0, y_0, vx_0, vy_0$ as in Section \ref{sec:results_schw}.
\\
As expected, we derive strong constraints on the enclosed mass within S2's orbit, $M_{encl,S2}$. The enclosed mass is compatible with zero, indicating that the orbits of the stars can be accurately described by Schwarzschild orbits without any extended mass surrounding the SMBH. In Figure \ref{fig:extmass_4s} we plot the $1 \sigma$ and $3 \sigma$ upper limits (corresponding to 68.3 \% c.i. and 99.7 \% c.i., see Appendix \ref{fit_details}) on this parameter, as a function of the power-law slope ($s$) for a power-law distribution, and as a function of the Plummer scale radius ($a$) for a Plummer distribution.
We find that we cannot distinguish between different density profiles, as a power-law with different slopes and a Plummer profile with different core radii fit the data equally well (we find no significant difference in $\chi^2$).
However, regardless of the specific density profile, we can reach a general conclusion.
The $1 \sigma$ upper limit on the enclosed mass within the orbit of S2 is lower than $\approx 1200 \, M_\odot$ for a power-law with slope $s \lesssim -1.2$ and for a Plummer profile with scale radius $a \lesssim 8$ mpc, which corresponds approximately to the apocenter distance of S2. 
These constraints represent a significant improvement over the results found in \cite{GRAVITY_2022}, where the $1 \sigma$ upper limit on any extended mass within S2's apocenter was found to be $\approx 3000 M_\odot$. In addition, this can be translated into an upper limit on the amount of retrograde precession induced by the mass distribution, which we find to be $\lesssim 1$ arcmin per orbit and is subdominant with respect to the prograde, Schwarzschild precession of $\sim 12$ arcmin per orbit. 
We emphasize that the data from S29, S38, and S55 are crucial for achieving such strong constraints (see Appendix \ref{fit_details} for a comparison with the results obtained from fitting the data of S2 only). \\

\subsection{Comparison with theoretical models for the stellar cusp}
\label{interpretation}
\begin{figure*}
   \centering
    \includegraphics[width=0.95\textwidth]{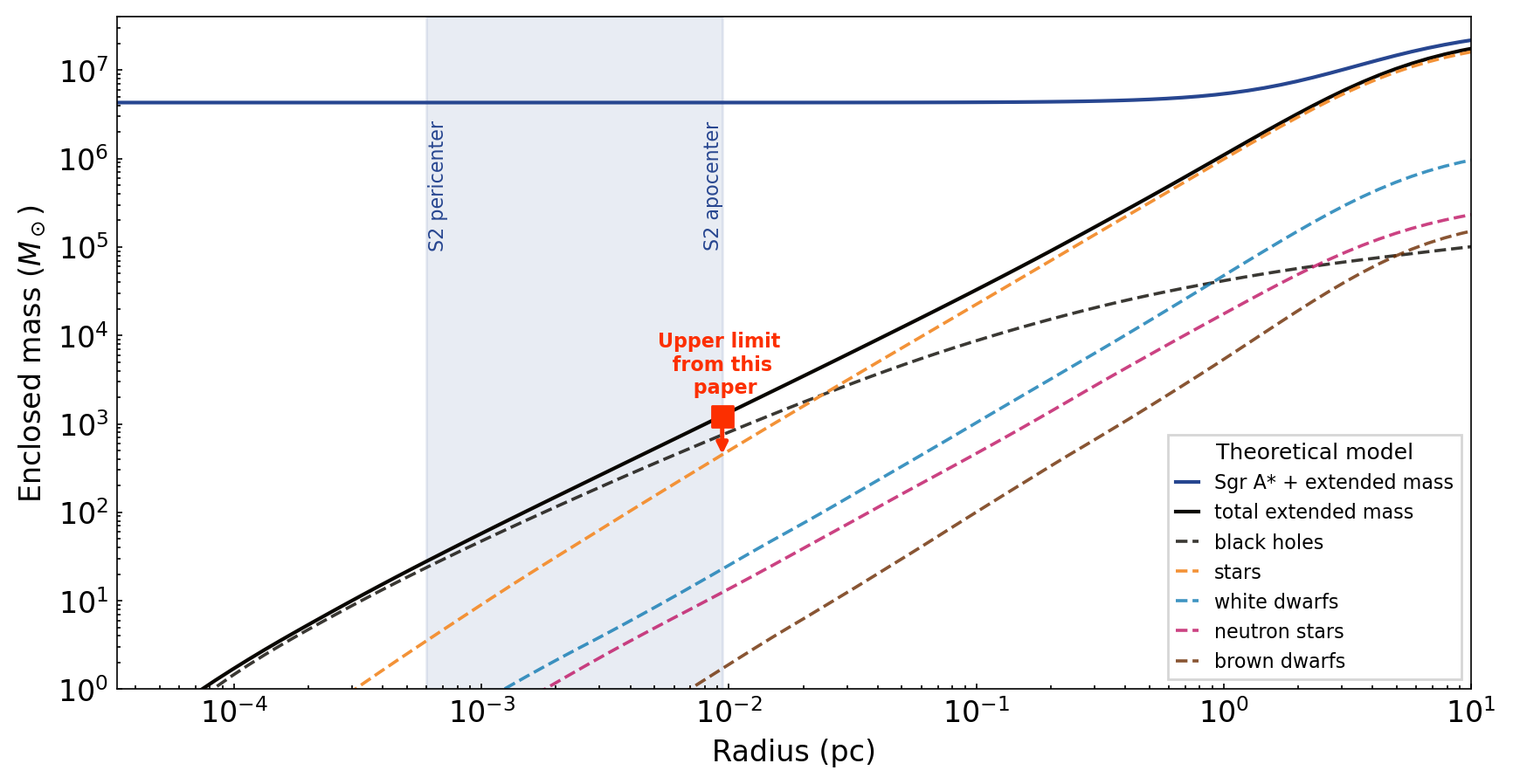}
   \caption{Enclosed mass in the GC as a function of radius derived from numerical simulations, using an updated version of the code from \cite{zhang_2023}, for a model which includes stellar black holes, neutron stars, white dwarfs, brown dwarfs and stars. The dashed colored lines show the predicted enclosed mass for each individual component, while the black solid line shows the total predicted enclosed mass considering all components. The predicted enclosed mass within S2's orbit is in agreement with our $1 \sigma$ upper limit of $\approx 1200 \, M_\odot$, indicated by the red square. The blue solid line shows the sum of the mass of Sgr A* to the total predicted enclosed mass. The blue shaded region illustrates the radial range of the orbit of S2.}
              \label{fig_reinhard}
    \end{figure*}

We now compare our derived upper limit on the enclosed mass within the orbit of S2 to what is predicted for a dynamically relaxed stellar cusp in the GC.
\cite{zhang_2023} presented a new Monte Carlo method, which allows to study the dynamical evolution of a star cluster with multiple mass components in the vicinity of a SMBH in a galactic nucleus. The code calculates the two-body relaxation process based on two-dimensional (energy and angular momentum) Fokker-Planck equations and includes the effects of the loss cone, giving thus a more realistic result that what can be obtained in the steady-state framework of \cite{BW_1976_1, BW_1977_2, Tal_2009}. 
Assuming a population of light objects with $m_l =1 \, M_\odot$ and a population of heavy remnants with $m_h = 10 \, M_\odot$, they find consistent results with the mass segregation solution described in \cite{Tal_2009,Preto_Pau_2010}. The heavy black holes sink towards the center due to dynamical friction and follow a steeper density profile as expected from mass segregation, reaching a slope between $\sim -2.3$ and $\sim -1.7$ in the inner regions. The light objects follow the expected slope of $-1.5$ if loss cone effects are ignored, while their density profile becomes shallower in the inner regions if the effects of the loss cone are included, reaching a slope of $ \sim -1.3$. \\
Here we use an updated version of the code (Zhang \& Amaro-Seoane, in prep.) which includes the potential from the stars, and not only that of the central SMBH, in the calculation of the two body relaxation of the energy and angular momentum of the particles in the simulation. This is particularly important for an accurate estimate of the density profile beyond the radius of influence of the SMBH, but also for the flux of particles into the loss cone.
We consider a model with five components, namely distinguishing between populations of stars of $m_{s}= 1 \, M_\odot$, brown dwarfs of $m_{bd}=0.05 \, M_\odot$, white dwarfs of $m_{wd}=0.6 \, M_\odot$, neutron stars of $m_{ns}=1.4 \, M_\odot$ and stellar black holes of $m_{bh}=10 \, M_\odot$, as in \cite{zhang_2023}. The fraction of brown dwarfs with respect to the stars at the beginning of the simulation is $f_{bd}=0.2$, of white dwarfs $f_{wd}=0.1$, of neutron stars $f_{ns}=0.01$ and of stellar black holes $f_{bh}=0.001$.
The initial density profile in the simulation is a Dehnen model \citep{Dehnen_1993} with $\gamma=1$, total mass $M_{cl}=3.2 \times 10^7 \, M_\odot$ and scale radius $r_a=1.86$ pc. We let the system evolve for one relaxation time and find that the radius at which the total enclosed mass in the simulation equals $2m_\bullet$ —the radius of influence of the SMBH \citep{Merritt+2013} — is $r = 3.9$ pc, consistent with findings from \cite{Chatzopoulos_2014, Feldmeier_2017}.
In Figure \ref{fig_reinhard} we show the enclosed mass profile resulting from the simulations as a function of distance from the SMBH, comparing it with our observational results. The enclosed mass within the orbit of S2 predicted by the simulations is $m_{encl,S2} = 1210 \, M_\odot$, which is in agreement with our upper limit $m_{encl,S2} \lesssim 1200 \, M_\odot$ for a power-law density profile with slope $s<-1.2$.  A similar result is obtained when considering a model with just two components, namely stars of $1 \, M_\odot$ and stellar black holes of $10 \, M_\odot$, giving a total enclosed mass within S2's orbit of $m_{encl,S2} = 1300 \, M_\odot$.\\
Since our upper limit is extremely close to the value predicted by simulations for a dynamically relaxed stellar cusp, we conclude that there is no substantial evidence for a significant enhancement of dark matter density near Sagittarius A*. However, it is important to note that, in principle, our orbital fits cannot distinguish between the mass contributions from the stellar cusp and a potential dark matter spike, as both are expected to follow similar power-law density distributions.\\
Moreover, the simulations indicate that stellar mass black holes dominate the mass distribution in this region, contributing significantly to the total mass enclosed within the orbit of S2. We find approximately 80 stellar mass black holes within S2's orbit, which is lower than the estimates provided by previous studies, such as \cite{Freitag_2006} and \cite{Tal_2009}. This finding suggests a downward trend in the predicted number of compact objects in this central region.

\subsection{Note on the granularity of the stellar cusp}
Our upper limit on the enclosed mass within S2's orbit of $m_{encl,S2} \lesssim 1200 \, M_\odot$ is low enough to raise questions about whether modeling the extended mass distribution as a smooth, spherically symmetric density profile is a valid approximation. 
Figure \ref{fig_reinhard} illustrates that, due to mass segregation, stellar black holes dominate the extended mass distribution within S2's orbit. Therefore, we consider a scenario in which a finite number of bodies generate the extended potential by distributing the total enclosed mass among $N$ objects of equal mass. This configuration can lead to deviations in the orbital motion of S-stars compared to a smooth density distribution, as spherical symmetry is broken and scattering events may occur. In this case, there is not only in-plane precession but also precession of the orbital plane, as discussed in \cite{Merritt_2010}. \\
An extreme case occurs when all the extended mass is concentrated in a single object, specifically an intermediate mass black hole (IMBH) companion to Sgr A*. It has been shown that an IMBH within the orbit of S2 can only have a mass $<10^3 \, M_\odot$, in order to be compatible with observations \citep{GRAVITY_2023a, Will_2023}. \\ 
We assume that a cluster of 100 point-mass particles, each with a mass of $10 \, M_\odot$, is distributed within the apocenter of S2, following a power-law density distribution of the form $\rho(r) \propto r^{-2}$. We further assume that these particles are fixed in space and do not interact with one another.
To assess the impact of this distribution on the orbit of S2, we conduct 100 mock observations of one full orbit of S2 around Sgr A* and the surrounding field objects, where each simulation corresponds to a different sampling of the field object positions to ensure adequate statistics. We then fit each simulated orbit for the $f_{SP}$ parameter (see Section \ref{sec:analysis_schw}), obtaining a distribution with a mean and median of $f_{SP}=0.94$, a standard deviation of 0.01, and a range of 0.09.
When repeating this experiment with a cluster of 20 point-mass particles, each of $50 \, M_\odot$, we obtain a distribution with a mean and median of $f_{SP}=0.95$, a standard deviation of 0.03, and a range of 0.23.\\ 
These results suggest that a distribution of stellar black holes around Sgr A* can in principle perturb our measurement of the Schwarzschild precession, causing a deviation in S2's orbit from the expected Schwarzschild orbit. This is an important effect to consider, especially as we achieve increasingly precise measurements of the Schwarzschild precession using GRAVITY data, which have led to an error on $f_{SP}$ of $\sim 0.1$ (Section \ref{sec:results_schw}). A detailed treatment of the effects of the granularity of the stellar cusp on stellar orbits is beyond the scope of this paper and will be addressed in detail in a forthcoming publication (Sadun Bordoni et al., in preparation).


\section{Conclusions}
\label{conclusions}
In this paper we have improved and extended the analysis conducted in \cite{GRAVITY_2020a, GRAVITY_2022}, adding the new interferometric data obtained with GRAVITY in 2022.
Particularly valuable, in addition to the well-known star S2, are data from the stars S29, S38, and S55. These stars have recently reached the pericenter of their orbits, allowing us to observe them around this critical phase with GRAVITY. \\
The orbital motions of S2, S29, S38 and S55 are perfectly compatible with Schwarzschild orbits around Sgr A* as predicted by GR, exhibiting prograde, in-plane precession of their pericenter angles. By performing a multi-star fit with this data, we detect the Schwarzschild precession of their orbits with a confidence level of approximately $\approx 10 \sigma$, marking a significant improvement over previous findings \citep{GRAVITY_2020a,GRAVITY_2022}.\\
We establish a stringent upper limit for the mass of any hypothetical extended mass distribution around Sgr A*, which would add a retrograde precession to the orbit of S2, counteracting the prograde, relativistic precession. This mass distribution could be composed of a dynamically relaxed cusp of old stars and stellar remnants and potentially of a dark matter spike.
We model it with a spherically symmetric density distribution, testing two plausible density profiles, namely a power-law and a Plummer profile. We find that, independently of the particular density profile, the enclosed mass within S2’s orbit is consistently compatible with zero. We set a strong upper limit of approximately $1200 \, M_\odot$ with a $1 \sigma$ confidence level, significantly improving upon the limits established in \cite{GRAVITY_2022}. 
Our findings align with theoretical predictions for a dynamically relaxed stellar cusp in the GC, composed of stars, brown dwarfs, white dwarfs, neutron stars, and stellar black holes, according to numerical simulations using an updated version of the code developed in \cite{zhang_2023}. This analysis predicts an enclosed mass of approximately $1210 \, M_\odot$ within S2’s orbit. Given that our upper limit is very close to this predicted value, we conclude that we find no evidence for a significant dark matter spike in the Galactic Center.\\
S2 is currently moving towards the apocenter of its orbit, which it will reach in 2026. We expect that GRAVITY data collected in the coming years, combined with ERIS spectroscopy, will further refine our constraints on the extended mass distribution in the GC,  as the mass distribution primarily influences stellar orbits in the apocenter half \citep{Heiss_2022}. This will allow refining the comparison with the theoretical predictions for the stellar cusp, which is of fundamental importance in order to understand the distribution of the faint, old main sequence stars and sub-giants in the GC. These stars are too faint to be currently detected with GRAVITY, but their detection could be in reach of future observations with the GRAVITY+ upgrade at the VLTI \citep{GRAVITY+} and the MICADO instrument at the ELT \citep{MICADO_ELT}. These stars could potentially be in tighter orbits around Sgr A* and could allow us to measure its spin and quadrupole moment.
Furthermore, the comparison between our observational constraints and theoretical predictions is also important to better understand the distribution of compact objects in the GC and in galactic nuclei in general. This could offer precious insights in view of the future LISA mission \citep{LISA_2017}, which will be able to detect the inspirals of compact objects into SMBHs (EMRIs) \citep{Amaro_Seoane_2007}. In fact, the rate of EMRIs depends strongly on the density distribution of compact remnants within $\sim 10$ mpc of the central SMBH \citep{Preto_Pau_2010}, which corresponds to the apocenter distance of S2 for the GC.

\begin{acknowledgements}
    We are very grateful to our funding agencies (MPG, ERC, CNRS [PNCG, PNGRAM], DFG, BMBF, Paris Observatory [CS, PhyFOG], Observatoire des Sciences de l’Univers de Grenoble, and the Fundação para a Ciência e Tecnologia), to ESO and the Paranal staff, and to the many scientific and technical staff members in our institutions, who helped to make NACO, SINFONI, and GRAVITY a reality. 
    JS is supported by the Deutsche Forschungsgemeinschaft (DFG, German Research Foundation) under Germany’s Excellence Strategy – EXC- 2094 – 390783311. A.A., A.F., P.G. and V.C. were supported by Fundação para a Ciência e a Tecnologia, with grants reference SFRH/BSAB/142940/2018, UIDB/00099/2020 and PTDC/FIS-AST/7002/2020.
    \end{acknowledgements}

\bibliography{bibliography_short_journals}

\begin{thebibliography}{60}
\expandafter\ifx\csname natexlab\endcsname\relax\def\natexlab#1{#1}\fi

\bibitem[{{Alexander}(2017)}]{Tal_2017}
{Alexander}, T. 2017, \href{http://dx.doi.org/10.1146/annurev-astro-091916-055306}{\color{magenta}\araa}, \href{https://ui.adsabs.harvard.edu/abs/2017ARA&A..55...17A}{55, 17}

\bibitem[{Alexander \& Hopman(2009)}]{Tal_2009}
Alexander, T. \& Hopman, C. 2009, \href{http://dx.doi.org/10.1088/0004-637x/697/2/1861}{\color{magenta}The Astrophysical Journal}, 697, 1861

\bibitem[{Amaro-Seoane {et~al.}(2017)Amaro-Seoane, Audley, Babak, Baker, Barausse, Bender, Berti, Binetruy, Born, Bortoluzzi, Camp, Caprini, Cardoso, Colpi, Conklin, Cornish, Cutler, Danzmann, Dolesi, Ferraioli, Ferroni, Fitzsimons, Gair, Bote, Giardini, Gibert, Grimani, Halloin, Heinzel, Hertog, Hewitson, Holley-Bockelmann, Hollington, Hueller, Inchauspe, Jetzer, Karnesis, Killow, Klein, Klipstein, Korsakova, Larson, Livas, Lloro, Man, Mance, Martino, Mateos, McKenzie, McWilliams, Miller, Mueller, Nardini, Nelemans, Nofrarias, Petiteau, Pivato, Plagnol, Porter, Reiche, Robertson, Robertson, Rossi, Russano, Schutz, Sesana, Shoemaker, Slutsky, Sopuerta, Sumner, Tamanini, Thorpe, Troebs, Vallisneri, Vecchio, Vetrugno, Vitale, Volonteri, Wanner, Ward, Wass, Weber, Ziemer, \& Zweifel}]{LISA_2017}
Amaro-Seoane, P., Audley, H., {et~al.} 2017, Laser Interferometer Space Antenna (ESA PUBLICATIONS DIVISION C/O ESTEC), submitted to ESA on January 13th in response to the call for missions for the L3 slot in the Cosmic Vision Programme

\bibitem[{Amaro-Seoane \& Chen(2014)}]{Amaro-Seoane_2014}
Amaro-Seoane, P. \& Chen, X. 2014, \href{http://dx.doi.org/10.1088/2041-8205/781/1/L18}{\color{magenta}The Astrophysical Journal Letters}, 781, L18

\bibitem[{Amaro-Seoane {et~al.}(2007)Amaro-Seoane, Gair, Freitag, Miller, Mandel, Cutler, \& Babak}]{Amaro_Seoane_2007}
Amaro-Seoane, P., Gair, J.~R., {et~al.} 2007, \href{http://dx.doi.org/10.1088/0264-9381/24/17/r01}{\color{magenta}Classical and Quantum Gravity}, 24, R113–R169

\bibitem[{Ang\'elil \& Saha(2014)}]{Ang_2014}
Ang\'elil, R. \& Saha, P. 2014, \href{http://dx.doi.org/10.1093/mnras/stu1686}{\color{magenta}Monthly Notices of the Royal Astronomical Society}, 444, 3780–3791

\bibitem[{Arg\"uelles {et~al.}(2019)Arg\"uelles, Krut, Rueda, \& Ruffini}]{arguelles_2019}
Arg\"uelles, C.~R., Krut, A., {et~al.} 2019, \href{http://dx.doi.org/10.1142/s021827181943003x}{\color{magenta}Int. J. Mod. Phys. D}, 28, 1943003

\bibitem[{{Bahcall} \& {Wolf}(1976)}]{BW_1976_1}
{Bahcall}, J.~N. \& {Wolf}, R.~A. 1976, \href{http://dx.doi.org/10.1086/154711}{\color{magenta}\apj}, \href{https://ui.adsabs.harvard.edu/abs/1976ApJ...209..214B}{209, 214}

\bibitem[{{Bahcall} \& {Wolf}(1977)}]{BW_1977_2}
{Bahcall}, J.~N. \& {Wolf}, R.~A. 1977, \href{http://dx.doi.org/10.1086/155534}{\color{magenta}\apj}, \href{https://ui.adsabs.harvard.edu/abs/1977ApJ...216..883B}{216, 883}

\bibitem[{{Bartko} {et~al.}(2010){Bartko}, {Martins}, {Trippe}, {Fritz}, {Genzel}, {Ott}, {Eisenhauer}, {Gillessen}, {Paumard}, {Alexander}, {Dodds-Eden}, {Gerhard}, {Levin}, {Mascetti}, {Nayakshin}, {Perets}, {Perrin}, {Pfuhl}, {Reid}, {Rouan}, {Zilka}, \& {Sternberg}}]{Barkto_2010}
{Bartko}, H., {Martins}, F., {et~al.} 2010, \href{http://dx.doi.org/10.1088/0004-637X/708/1/834}{\color{magenta}\apj}, \href{https://ui.adsabs.harvard.edu/abs/2010ApJ...708..834B}{708, 834}

\bibitem[{{Baumgardt} {et~al.}(2018){Baumgardt}, {Amaro-Seoane}, \& {Sch{\"o}del}}]{Baumb_2018}
{Baumgardt}, H., {Amaro-Seoane}, P., \& {Sch{\"o}del}, R. 2018, \href{http://dx.doi.org/10.1051/0004-6361/201730462}{\color{magenta}\aap}, \href{https://ui.adsabs.harvard.edu/abs/2018A&A...609A..28B}{609, A28}

\bibitem[{Becerra-Vergara {et~al.}(2020)Becerra-Vergara, Arg\"uelles, Krut, Rueda, \& Ruffini}]{Becerra_Vergara_2020}
Becerra-Vergara, E.~A., Arg\"uelles, C.~R., {et~al.} 2020, \href{http://dx.doi.org/10.1051/0004-6361/201935990}{\color{magenta}\aap}, 641, A34

\bibitem[{{Buchholz, R. M.} {et~al.}(2009){Buchholz, R. M.}, {Sch\"odel, R.}, \& {Eckart, A.}}]{Buchholz_2009}
{Buchholz, R. M.}, {Sch\"odel, R.}, \& {Eckart, A.} 2009, \href{http://dx.doi.org/10.1051/0004-6361/200811497}{\color{magenta}A\&A}, 499, 483

\bibitem[{Capuzzo-Dolcetta \& Sadun-Bordoni(2023)}]{Sad_2023}
Capuzzo-Dolcetta, R. \& Sadun-Bordoni, M. 2023, \href{http://dx.doi.org/10.1093/mnras/stad1317}{\color{magenta}Monthly Notices of the Royal Astronomical Society}, 522, 5828

\bibitem[{Chatzopoulos {et~al.}(2014)Chatzopoulos, Fritz, Gerhard, Gillessen, Wegg, Genzel, \& Pfuhl}]{Chatzopoulos_2014}
Chatzopoulos, S., Fritz, T.~K., {et~al.} 2014, \href{http://dx.doi.org/10.1093/mnras/stu2452}{\color{magenta}Monthly Notices of the Royal Astronomical Society}, 447, 948–968

\bibitem[{{Davies} {et~al.}(2018){Davies}, {Alves}, {Cl{\'e}net}, {Lang-Bardl}, {Nicklas}, {Pott}, {Ragazzoni}, {Tolstoy}, {Amico}, {Anwand-Heerwart}, {Barboza}, {Barl}, {Baudoz}, {Bender}, {Bezawada}, {Bizenberger}, {Boland}, {Bonifacio}, {Borgo}, {Buey}, {Chapron}, {Chemla}, {Cohen}, {Czoske}, {D{\'e}o}, {Disseau}, {Dreizler}, {Dupuis}, {Fabricius}, {Falomo}, {Fedou}, {F{\"o}rster Schreiber}, {Garrel}, {Geis}, {Gemperlein}, {Gendron}, {Genzel}, {Gillessen}, {Gl{\"u}ck}, {Grupp}, {Hartl}, {H{\"a}user}, {Hess}, {Hofferbert}, {Hopp}, {H{\"o}rmann}, {Hubert}, {Huby}, {Huet}, {Hutterer}, {Ives}, {Janssen}, {Jellema}, {Kausch}, {Kerber}, {Kravcar}, {Le Ruyet}, {Leschinski}, {Mandla}, {Manhart}, {Massari}, {Mei}, {Merlin}, {Mohr}, {Monna}, {Muench}, {M{\"u}ller}, {Musters}, {Navarro}, {Neumann}, {Neumayer}, {Niebsch}, {Plattner}, {Przybilla}, {Rabien}, {Ramlau}, {Ramos}, {Ramsay}, {Rhode}, {Richter}, {Richter}, {Rix}, {Rodeghiero}, {Rohloff}, {Rosensteiner}, {Rousset}, {Schlichter}, {Schubert}, {Sevin}, {Stuik},
  {Sturm}, {Thomas}, {Tromp}, {Verdoes-Kleijn}, {Vidal}, {Wagner}, {Wegner}, {Zeilinger}, {Ziegleder}, {Ziegler}, \& {Zins}}]{MICADO_ELT}
{Davies}, R., {Alves}, J., {et~al.} 2018, in Society of Photo-Optical Instrumentation Engineers (SPIE) Conference Series, Vol. 10702, Ground-based and Airborne Instrumentation for Astronomy VII, ed. C.~J. {Evans}, L.~{Simard}, \& H.~{Takami}, \href{https://ui.adsabs.harvard.edu/abs/2018SPIE10702E..1SD}{107021S}

\bibitem[{{Davies, R.} {et~al.}(2023){Davies, R.}, {Absil, O.}, {Agapito, G.}, {Agudo Berbel, A.}, {Baruffolo, A.}, {Biliotti, V.}, {Black, M.}, {Bonaglia, M.}, {Bonse, M.}, {Briguglio, R.}, {Campana, P.}, {Cao, Y.}, {Carbonaro, L.}, {Cortes, A.}, {Cresci, G.}, {Dallilar, Y.}, {Dannert, F.}, {De Rosa, R. J.}, {Deysenroth, M.}, {Di Antonio, I.}, {Di Cianno, A.}, {Di Rico, G.}, {Doelman, D.}, {Dolci, M.}, {Dorn, R.}, {Eisenhauer, F.}, {Esposito, S.}, {Fantinel, D.}, {Ferruzzi, D.}, {Feuchtgruber, H.}, {Finger, G.}, {F\"orster Schreiber, N. M.}, {Gao, X.}, {Gemperlein, H.}, {Genzel, R.}, {Gillessen, S.}, {Ginski, C.}, {Glauser, A. M.}, {Glindemann, A.}, {Grani, P.}, {Hartl, M.}, {Hayoz, J.}, {Heida, M.}, {Henry, D.}, {Hofmann, R.}, {Huber, H.}, {Kasper, M.}, {Keller, C.}, {Kenworthy, M.}, {Kravchenko, K.}, {Kuntschner, H.}, {Lacour, S.}, {Lightfoot, J.}, {Lunney, D.}, {Lutz, D.}, {Macintosh, M.}, {Mannucci, F.}, {Marsset, M.}, {Modigliani, A.}, {Neeser, M.}, {Orban de Xivry, G.}, {Ott, T.}, {Pallanca, L.},
  {Patapis, P.}, {Pearson, D.}, {Pe\~na, E.}, {Percheron, I.}, {Puglisi, A.}, {Quanz, S. P.}, {Rabien, S.}, {Rau, C.}, {Riccardi, A.}, {Salasnich, B.}, {Schmid, H.-M.}, {Schubert, J.}, {Serra, B.}, {Shimizu, T.}, {Snik, F.}, {Sturm, E.}, {Tacconi, L.}, {Taylor, W.}, {Valentini, A.}, {Waring, C.}, {Wiezorrek, E.}, \& {Xompero, M.}}]{ERIS_2023}
{Davies, R.}, {Absil, O.}, {et~al.} 2023, \href{http://dx.doi.org/10.1051/0004-6361/202346559}{\color{magenta}A\&A}, 674, A207

\bibitem[{{Dehnen}(1993)}]{Dehnen_1993}
{Dehnen}, W. 1993, \href{http://dx.doi.org/10.1093/mnras/265.1.250}{\color{magenta}\mnras}, \href{https://ui.adsabs.harvard.edu/abs/1993MNRAS.265..250D}{265, 250}

\bibitem[{Do {et~al.}(2009)Do, Ghez, Morris, Lu, Matthews, Yelda, \& Larkin}]{Do_2009}
Do, T., Ghez, A.~M., {et~al.} 2009, \href{http://dx.doi.org/10.1088/0004-637X/703/2/1323}{\color{magenta}The Astrophysical Journal}, 703, 1323

\bibitem[{Do {et~al.}(2019)Do, Hees, Ghez, Martinez, Chu, Jia, Sakai, Lu, Gautam, O'Neil, Becklin, Morris, Matthews, Nishiyama, Campbell, Chappell, Chen, Ciurlo, Dehghanfar, Gallego-Cano, Kerzendorf, Lyke, Naoz, Saida, Sch\"odel, Takahashi, Takamori, Witzel, \& Wizinowich}]{Do_2019a}
Do, T., Hees, A., {et~al.} 2019, \href{http://dx.doi.org/10.1126/science.aav8137}{\color{magenta}Science}, 365, 664

\bibitem[{Eisenhauer {et~al.}(2005)Eisenhauer, Genzel, Alexander, Abuter, Paumard, Ott, Gilbert, Gillessen, Horrobin, Trippe, Bonnet, Dumas, Hubin, Kaufer, Kissler-Patig, Monnet, Strobele, Szeifert, Eckart, Schodel, \& Zucker}]{Eisenhauer_2005}
Eisenhauer, F., Genzel, R., {et~al.} 2005, \href{http://dx.doi.org/10.1086/430667}{\color{magenta}\apj}, 628, 246

\bibitem[{{Event Horizon Telescope Collaboration}(2022)}]{EHT_2022}
{Event Horizon Telescope Collaboration}. 2022, \href{http://dx.doi.org/10.3847/2041-8213/ac6674}{\color{magenta}\apj}, 930, L12

\bibitem[{Feldmeier-Krause {et~al.}(2016)Feldmeier-Krause, Zhu, Neumayer, van~de Ven, de~Zeeuw, \& Sch\"odel}]{Feldmeier_2017}
Feldmeier-Krause, A., Zhu, L., {et~al.} 2016, \href{http://dx.doi.org/10.1093/mnras/stw3377}{\color{magenta}Monthly Notices of the Royal Astronomical Society}, 466, 4040

\bibitem[{Foschi {et~al.}(2023)Foschi, Abuter, Aimar, Amaro~Seoane, Amorim, Baub\"ock, Berger, Bonnet, Bourdarot, Brandner, Cardoso, Cl\'enet, Dallilar, Davies, de~Zeeuw, Defr\`ere, Dexter, Drescher, Eckart, Eisenhauer, Ferreira, F\"orster~Schreiber, Garcia, Gao, Gendron, Genzel, Gillessen, Gomes, Habibi, Haubois, Heißel, Henning, Hippler, H\"onig, Horrobin, Jochum, Jocou, Kaufer, Kervella, Kreidberg, Lacour, Lapeyr\`ere, Le~Bouquin, L\'ena, Lutz, Millour, Ott, Paumard, Perraut, Perrin, Pfuhl, Rabien, Ribeiro, Sadun~Bordoni, Scheithauer, Shangguan, Shimizu, Stadler, Straub, Straubmeier, Sturm, Sykes, Tacconi, Vincent, von Fellenberg, Widmann, Wieprecht, Wiezorrek, Woillez, \& Collaboration}]{GRAVITY_2023_c}
Foschi, A., Abuter, R., {et~al.} 2023, \href{http://dx.doi.org/10.1093/mnras/stad1939}{\color{magenta}Monthly Notices of the Royal Astronomical Society}, 524, 1075

\bibitem[{{Frank} \& {Rees}(1976)}]{frank_rees_1976}
{Frank}, J. \& {Rees}, M.~J. 1976, \href{http://dx.doi.org/10.1093/mnras/176.3.633}{\color{magenta}\mnras}, \href{https://ui.adsabs.harvard.edu/abs/1976MNRAS.176..633F}{176, 633}

\bibitem[{Freitag {et~al.}(2006)Freitag, Amaro-Seoane, \& Kalogera}]{Freitag_2006}
Freitag, M., Amaro-Seoane, P., \& Kalogera, V. 2006, \href{http://dx.doi.org/10.1086/506193}{\color{magenta}The Astrophysical Journal}, 649, 91–117

\bibitem[{{Gallego-Cano} {et~al.}(2018){Gallego-Cano}, {Sch{\"o}del}, {Dong}, {Nogueras-Lara}, {Gallego-Calvente}, {Amaro-Seoane}, \& {Baumgardt}}]{Schodel_2018_1}
{Gallego-Cano}, E., {Sch{\"o}del}, R., {et~al.} 2018, \href{http://dx.doi.org/10.1051/0004-6361/201730451}{\color{magenta}\aap}, \href{https://ui.adsabs.harvard.edu/abs/2018A&A...609A..26G}{609, A26}

\bibitem[{Genzel {et~al.}(2010)Genzel, Eisenhauer, \& Gillessen}]{Genzel_2010}
Genzel, R., Eisenhauer, F., \& Gillessen, S. 2010, \href{http://dx.doi.org/10.1103/revmodphys.82.3121}{\color{magenta}Reviews of Modern Physics}, 82, 3121–3195

\bibitem[{{Ghez} {et~al.}(2003){Ghez}, {Duch{\^e}ne}, {Matthews}, {Hornstein}, {Tanner}, {Larkin}, {Morris}, {Becklin}, {Salim}, {Kremenek}, {Thompson}, {Soifer}, {Neugebauer}, \& {McLean}}]{Ghez_2003}
{Ghez}, A.~M., {Duch{\^e}ne}, G., {et~al.} 2003, \href{http://dx.doi.org/10.1086/374804}{\color{magenta}\apjl}, \href{https://ui.adsabs.harvard.edu/abs/2003ApJ...586L.127G}{586, L127}

\bibitem[{{Ghez} {et~al.}(2008){Ghez}, {Salim}, {Weinberg}, {Lu}, {Do}, {Dunn}, {Matthews}, {Morris}, {Yelda}, {Becklin}, {Kremenek}, {Milosavljevic}, \& {Naiman}}]{Ghez_2008}
{Ghez}, A.~M., {Salim}, S., {et~al.} 2008, \href{http://dx.doi.org/10.1086/592738}{\color{magenta}\apj}, \href{https://ui.adsabs.harvard.edu/abs/2008ApJ...689.1044G}{689, 1044}

\bibitem[{Gillessen {et~al.}(2017)Gillessen, Plewa, Eisenhauer, Sari, Waisberg, Habibi, Pfuhl, George, Dexter, von Fellenberg, Ott, \& Genzel}]{Gillessen_2017}
Gillessen, S., Plewa, P.~M., {et~al.} 2017, \href{http://dx.doi.org/10.3847/1538-4357/aa5c41}{\color{magenta}\apj}, 837, 30

\bibitem[{{Gondolo} \& {Silk}(1999)}]{Silk_99}
{Gondolo}, P. \& {Silk}, J. 1999, \href{http://dx.doi.org/10.1103/PhysRevLett.83.1719}{\color{magenta}\prl}, \href{https://ui.adsabs.harvard.edu/abs/1999PhRvL..83.1719G}{83, 1719}

\bibitem[{{GRAVITY Collaboration}(2017)}]{GRAVITY_2017}
{GRAVITY Collaboration}. 2017, \href{http://dx.doi.org/10.1051/0004-6361/201730838}{\color{magenta}\aap}, 602, A94

\bibitem[{{GRAVITY Collaboration}(2018)}]{GRAVITY_2018b}
{GRAVITY Collaboration}. 2018, \href{http://dx.doi.org/10.1051/0004-6361/201834294}{\color{magenta}\aap}, 618, L10

\bibitem[{{GRAVITY Collaboration}(2018a)}]{GRAVITY_2018}
{GRAVITY Collaboration}. 2018a, \href{http://dx.doi.org/10.1051/0004-6361/201833718}{\color{magenta}\aap}, 615, L15

\bibitem[{{GRAVITY Collaboration}(2019)}]{GRAVITY_2019}
{GRAVITY Collaboration}. 2019, \href{http://dx.doi.org/10.1051/0004-6361/201935656}{\color{magenta}A\&A}, 625, L10

\bibitem[{{GRAVITY Collaboration}(2020)}]{GRAVITY_2020a}
{GRAVITY Collaboration}. 2020, \href{http://dx.doi.org/10.1051/0004-6361/202037813}{\color{magenta}\aap}, 636, L5

\bibitem[{{GRAVITY Collaboration}(2022)}]{GRAVITY_2022}
{GRAVITY Collaboration}. 2022, \href{http://dx.doi.org/10.1051/0004-6361/202142465}{\color{magenta}\aap}, 657, L12

\bibitem[{{GRAVITY Collaboration}(2023{\natexlab{a}})}]{GRAVITY_2023_flares}
{GRAVITY Collaboration}. 2023{\natexlab{a}}, \href{http://dx.doi.org/10.1051/0004-6361/202347416}{\color{magenta}A\&A}, 677, L10

\bibitem[{{GRAVITY Collaboration}(2023{\natexlab{b}})}]{GRAVITY_2023a}
{GRAVITY Collaboration}. 2023{\natexlab{b}}, \href{http://dx.doi.org/10.1051/0004-6361/202245132}{\color{magenta}A\&A}, 672, A63

\bibitem[{{GRAVITY Collaboration}(2024)}]{GRAVITY_2024}
{GRAVITY Collaboration}. 2024, \href{http://dx.doi.org/10.1093/mnras/stae423}{\color{magenta}Monthly Notices of the Royal Astronomical Society}, 530, 3740

\bibitem[{{GRAVITY+ Collaboration} {et~al.}(2022){GRAVITY+ Collaboration}, Abuter, Alarcon, Allouche, Amorim, Bailet, Bedigan, Berdeu, Berger, Berio, Bigioli, Blaho, Boebion, Bolzer, Bonnet, Bourdarot, Bourget, Brandner, Cardenas, Conzelmann, Comin, Cl\'enet, Courtney-Barrer, Dallilar, Davies, Defr\`ere, Delboulb\'e, Delplancke-Str\"obele, Dembet, De~Zeeuw, Drescher, Eckart, Édouard, Eisenhauer, Fabricius, Feuchtgruber, Finger, F\"orster~Schreiber, Fuenteseca, Garcia, Garcia, Gao, Gendron, Genzel, Gil, Gillessen, Gomes, Gont\'e, Gouvret, Guajardo, Guidolin, Guieu, Guzmann, Hackenberg, Haddad, Hartl, Haubois, Haußmann, Heißel, Henning, Hippler, H\"onig, Horrobin, Hubin, Jacqmart, Jocou, Kaufer, Kervella, Kirchbauer, Kolb, Korhonen, Kreidberg, Krempl, Lacour, Lagarde, Lai, Lapeyr\`ere, Laugier, Le~Bouquin, Leftley, L\'ena, Lewis, Lutz, Magnard, Mang, Marcotto, Maurel, M\'erand, Millour, More, Nowacki, Nowak, Oberti, Olivares, Ott, Pallanca, Paumard, Perraut, Perrin, Petrov, Pfuhl, Pourr\'e, Rabien, Rau,
  Riquelme, Robbe-Dubois, Rochat, Salman, Scherbarth, Sch\"oller, Schubert, Schuhler, Shangguan, Shchekaturov, Shimizu, Scheithauer, Sevin, Soenke, Soulez, Spang, Stadler, Straubmeier, Sturm, Sykes, Tacconi, Tischer, Tristram, Vincent, Von~Fellenberg, Uysal, Widmann, Wieprecht, Wiezorrek, Woillez, Yaz\i~c\i, \& Zins}]{GRAVITY+}
{GRAVITY+ Collaboration}, Abuter, R., {et~al.} 2022, \href{http://dx.doi.org/10.18727/0722-6691/5285}{\color{magenta}Published in The Messenger vol. 189}, pp. 17-22, December 2022.

\bibitem[{Habibi {et~al.}(2017)Habibi, Gillessen, Martins, Eisenhauer, Plewa, Pfuhl, George, Dexter, Waisberg, Ott, Fellenberg, Baub\"ock, Jimenez-Rosales, \& Genzel}]{Habibi_2017}
Habibi, M., Gillessen, S., {et~al.} 2017, \href{http://dx.doi.org/10.3847/1538-4357/aa876f}{\color{magenta}The Astrophysical Journal}, 847, 120

\bibitem[{{Heißel, G.} {et~al.}(2022){Heißel, G.}, {Paumard, T.}, {Perrin, G.}, \& {Vincent, F.}}]{Heiss_2022}
{Heißel, G.}, {Paumard, T.}, {et~al.} 2022, \href{http://dx.doi.org/10.1051/0004-6361/202142114}{\color{magenta}A\&A}, 660, A13

\bibitem[{Linial \& Sari(2022)}]{Linial_2022}
Linial, I. \& Sari, R. 2022, \href{http://dx.doi.org/10.3847/1538-4357/ac9bfd}{\color{magenta}The Astrophysical Journal}, 940, 101

\bibitem[{Merritt(2013)}]{Merritt+2013}
Merritt, D. 2013, Dynamics and Evolution of Galactic Nuclei (Princeton: Princeton University Press)

\bibitem[{Merritt {et~al.}(2010)Merritt, Alexander, Mikkola, \& Will}]{Merritt_2010}
Merritt, D., Alexander, T., {et~al.} 2010, \href{http://dx.doi.org/10.1103/physrevd.81.062002}{\color{magenta}Physical Review D}, 81

\bibitem[{{Peebles}(1972)}]{Peebles_1972}
{Peebles}, P.~J.~E. 1972, \href{http://dx.doi.org/10.1086/151797}{\color{magenta}\apj}, \href{https://ui.adsabs.harvard.edu/abs/1972ApJ...178..371P}{178, 371}

\bibitem[{{Plewa} {et~al.}(2015){Plewa}, {Gillessen}, {Eisenhauer}, {Ott}, {Pfuhl}, {George}, {Dexter}, {Habibi}, {Genzel}, {Reid}, \& {Menten}}]{Plewa_2015}
{Plewa}, P.~M., {Gillessen}, S., {et~al.} 2015, \href{http://dx.doi.org/10.1093/mnras/stv1910}{\color{magenta}\mnras}, \href{https://ui.adsabs.harvard.edu/abs/2015MNRAS.453.3234P}{453, 3234}

\bibitem[{{Plummer}(1911)}]{Plummer_1911}
{Plummer}, H.~C. 1911, \href{http://dx.doi.org/10.1093/mnras/71.5.460}{\color{magenta}\mnras}, \href{https://ui.adsabs.harvard.edu/abs/1911MNRAS..71..460P}{71, 460}

\bibitem[{{Preto} \& {Amaro-Seoane}(2010)}]{Preto_Pau_2010}
{Preto}, M. \& {Amaro-Seoane}, P. 2010, \href{http://dx.doi.org/10.1088/2041-8205/708/1/L42}{\color{magenta}\apjl}, \href{https://ui.adsabs.harvard.edu/abs/2010ApJ...708L..42P}{708, L42}

\bibitem[{Rose \& MacLeod(2024)}]{rose_2023}
Rose, S.~C. \& MacLeod, M. 2024, \href{http://dx.doi.org/10.3847/2041-8213/ad251f}{\color{magenta}The Astrophysical Journal Letters}, 963, L17

\bibitem[{{Sch{\"o}del} {et~al.}(2018){Sch{\"o}del}, {Gallego-Cano}, {Dong}, {Nogueras-Lara}, {Gallego-Calvente}, {Amaro-Seoane}, \& {Baumgardt}}]{Schodel_2018_2}
{Sch{\"o}del}, R., {Gallego-Cano}, E., {et~al.} 2018, \href{http://dx.doi.org/10.1051/0004-6361/201730452}{\color{magenta}\aap}, \href{https://ui.adsabs.harvard.edu/abs/2018A&A...609A..27S}{609, A27}

\bibitem[{{Sch{\"o}del} {et~al.}(2002){Sch{\"o}del}, {Ott}, {Genzel}, {Hofmann}, {Lehnert}, {Eckart}, {Mouawad}, {Alexander}, {Reid}, {Lenzen}, {Hartung}, {Lacombe}, {Rouan}, {Gendron}, {Rousset}, {Lagrange}, {Brandner}, {Ageorges}, {Lidman}, {Moorwood}, {Spyromilio}, {Hubin}, \& {Menten}}]{Schodel_2002}
{Sch{\"o}del}, R., {Ott}, T., {et~al.} 2002, \href{http://dx.doi.org/10.1038/nature01121}{\color{magenta}\nat}, \href{https://ui.adsabs.harvard.edu/abs/2002Natur.419..694S}{419, 694}

\bibitem[{Shen {et~al.}(2024)Shen, Yuan, Jiang, Tsai, Yuan, \& Fan}]{Shen_2023}
Shen, Z.-Q., Yuan, G.-W., {et~al.} 2024, \href{http://dx.doi.org/10.1093/mnras/stad3282}{\color{magenta}Mon. Not. Roy. Astron. Soc.}, 527, 3196

\bibitem[{Viollier {et~al.}(1993)Viollier, Trautmann, \& Tupper}]{viollier_1993}
Viollier, R., Trautmann, D., \& Tupper, G. 1993, \href{http://dx.doi.org/https://doi.org/10.1016/0370-2693(93)91141-9}{\color{magenta}Physics Letters B}, 306, 79

\bibitem[{{Waisberg} {et~al.}(2018){Waisberg}, {Dexter}, {Gillessen}, {Pfuhl}, {Eisenhauer}, {Plewa}, {Baub{\"o}ck}, {Jimenez-Rosales}, {Habibi}, {Ott}, {von Fellenberg}, {Gao}, {Widmann}, \& {Genzel}}]{Waisb_2018}
{Waisberg}, I., {Dexter}, J., {et~al.} 2018, \href{http://dx.doi.org/10.1093/mnras/sty476}{\color{magenta}\mnras}, \href{https://ui.adsabs.harvard.edu/abs/2018MNRAS.476.3600W}{476, 3600}

\bibitem[{Will(1993)}]{Will_1993}
Will, C.~M. 1993, Theory and Experiment in Gravitational Physics (Cambridge University Press)

\bibitem[{Will {et~al.}(2023)Will, Naoz, Hees, Tucker, Zhang, Do, \& Ghez}]{Will_2023}
Will, C.~M., Naoz, S., {et~al.} 2023, \href{http://dx.doi.org/10.3847/1538-4357/ad09b3}{\color{magenta}The Astrophysical Journal}, 959, 58

\bibitem[{Zhang \& Amaro-Seoane(2024)}]{zhang_2023}
Zhang, F. \& Amaro-Seoane, P. 2024, \href{http://dx.doi.org/10.3847/1538-4357/ad0f1a}{\color{magenta}The Astrophysical Journal}, 961, 232

\end{thebibliography}

\begin{appendix}

\section{Predicting the improvement on the SP detection with future observations}
\label{sp_prediction}
In order to predict how much we will be able to improve our constraint on the $f_{SP}$ parameter by continuing to monitor the orbit of S2 with GRAVITY and ERIS, we carried out mock observations assuming that we will get, between March and September of each year (GC observing season from Paranal observatory): 
\begin{itemize}
    \item 10 GRAVITY data points per year with $50 \, \mu as$ astrometric accuracy,
    \item 3 radial velocity data points per year with ERIS with $10 \, km/s$ accuracy.
\end{itemize} 

\begin{figure}
   \centering
    \includegraphics[width=0.5\textwidth]{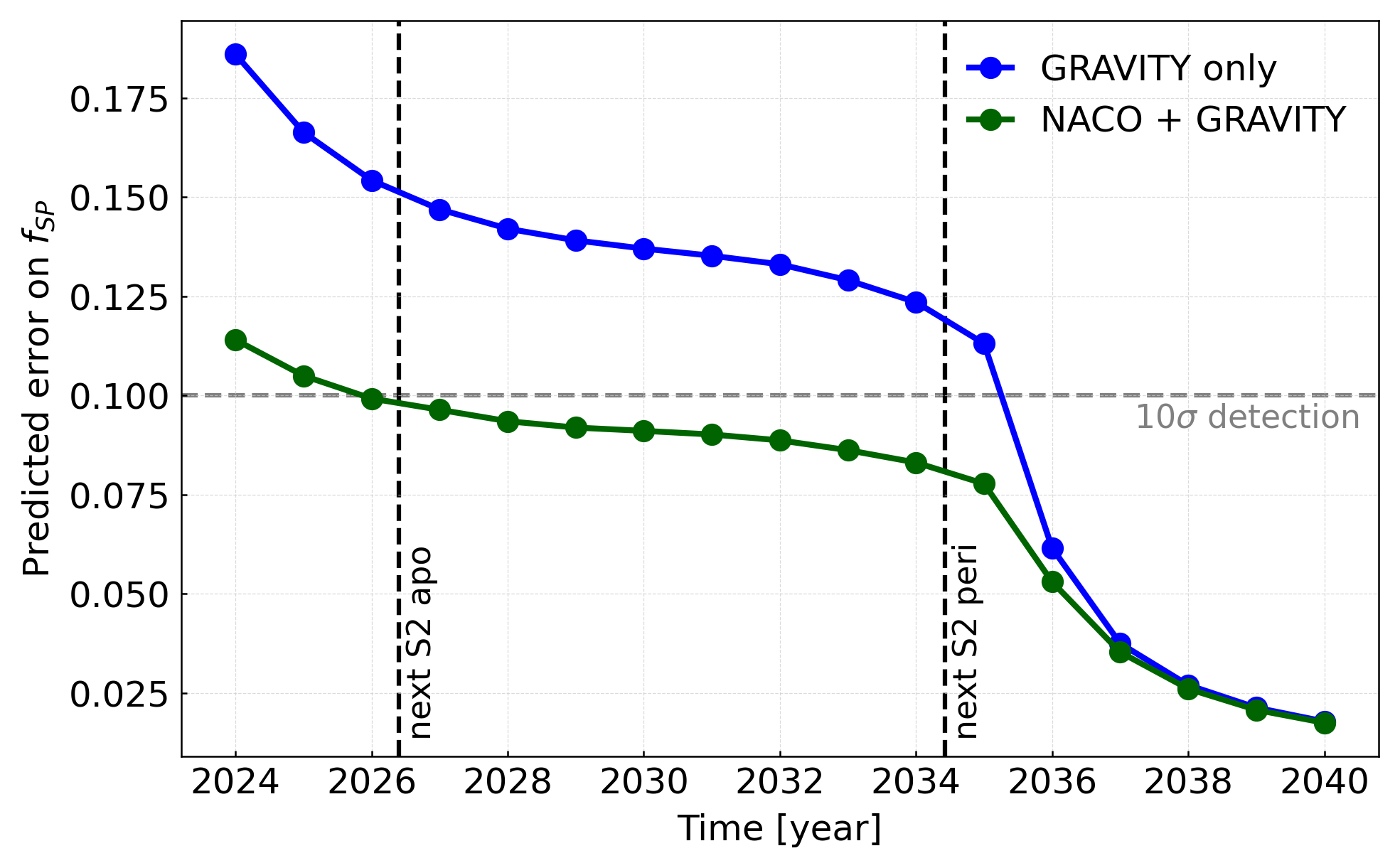}
   \caption{Predicted error on $f_{SP}$ as a function of time, obtained through mock observations of S2 with GRAVITY and ERIS. The blue points show the result keeping only GRAVITY data in the fit, namely excluding the NACO data and removing the reference frame parameters. The green points show the result keeping both NACO and GRAVITY data. 
   }
              \label{fig_simu}
    \end{figure}

We then fit the actual S2 data set together with the mock data set and obtain a prediction on the error on $f_{SP}$ in function of time.
The result of this analysis is shown in Figure \ref{fig_simu}. We compare the case in which we keep the NACO data in the fit with a prior on the reference frame parameters, with the case in which we use GRAVITY data only and we thus remove the reference frame parameters. 
This shows that GRAVITY data taken in the following years, until the S2 star will reach the next apocenter in 2026, will only moderately improve the SP detection. The predicted error remains then almost constant until $\approx 2034$, namely the time of the next pericenter passage of S2, and starts then to rapidly decrease. This is not surprising, because the Schwarzschild precession happens mostly around pericenter \citep{Ang_2014, GRAVITY_2020a, Heiss_2022}.\\
This analysis also tells us how valuable the NACO imaging data will still be in the near future, at least until S2 will reach the next pericenter passage in 2034. In fact, Figure \ref{fig_simu} shows that the constraint on $f_{SP}$ will be stronger including NACO data in the orbital fit until a few years after the next pericenter passage of S2. In order to reach the same accuracy that we will be able to get on the SP detection keeping the NACO data in the fit, but using only GRAVITY data and eliminating the NACO zero points $x_0, y_0, vx_0, vy_0$, we would need to wait until $\approx 2037$, when we will have observed two consecutive pericenter passages with GRAVITY.

\section{Fit details}
\label{fit_details}

\begin{table}[h]
\centering
\begin{tabular}{|ccccc|}
  \hline
   parameter & value & error & prior & error \\
  \hline
  $M_\bullet$ [$10^6 M_\odot$]& 4.2996 &	0.0118&-&-  \\

  $R_0$ [pc]& 8275.9	& 8.6 &-&- \\

  $x_0$ [mas]& -0.72370	& 0.08707 & -7.79 $\times 10^{-4}$& 7.38 $\times 10^{-5}$ \\

  $y_0$ [mas] & -0.11921 &	0.07391 & -1.18$\times 10^{-4}$& 7.93 $\times 10^{-5}$\\

  $vx_0$ [mas/yr] & 0.078241 & 0.005728 & 6.64$\times 10^{-5}$	& 7.27$\times 10^{-6}$\\

  $vy_0$ [mas/yr] & 0.040542	& 0.005620 & 3.70$\times 10^{-5}$ &	7.76$\times 10^{-6}$\\

  $vz_0$ [km/s] & -1.9983	& 1.3325 & 0& 5\\

  $f_{SP}$ & 1.1350 &	0.1104 &-&- \\

  \hline \hline

  \multicolumn{5}{|c|}{\rule{0pt}{2ex}\textbf{S2}}\\
  \hline
  $a$ [as] & 0.12502& 3$\times 10^{-5}$ &-&- \\

  $e$  & 0.88444 & 6$\times 10^{-5}$ &-&- \\

  $i$ [$^{\circ}$] & 134.67& 0.02 &-&- \\

  $\Omega$ [$^{\circ}$] & 228.21& 0.03 &-&- \\

  $\omega$ [$^{\circ}$] & 66.279 & 0.029 &-&- \\

  $t_{peri}$ [yr] & 2018.3789 & 1$\times 10^{-4}$ &-&- \\

    \hline \hline

  \multicolumn{5}{|c|}{\rule{0pt}{2ex}\textbf{S29}} \\
  \hline

  $a$ [as] & 0.39025 & 9.4$\times 10^{-4}$ &-&- \\

  $e$  & 0.96880 & 9$\times 10^{-5}$ &-&- \\

  $i$ [$^{\circ}$] & 144.24 & 0.09 &-&- \\

  $\Omega$ [$^{\circ}$] & 4.9259 & 0.1590 &-&- \\

  $\omega$ [$^{\circ}$] & 203.68 & 0.17 &-&- \\

  $t_{peri}$ [yr] & 2021.4102 & $\times 10^{-4}$ &-&- \\

    \hline \hline

   \multicolumn{5}{|c|}{\rule{0pt}{2ex}\textbf{S38}}\\
   \hline

  $a$ [as] & 0.14249 & 4$\times 10^{-5}$ &-&- \\

  $e$  & 0.81807 & 2.2$\times 10^{-4}$ &-&- \\

  $i$ [$^{\circ}$] & 168.69 & 0.19 &-&- \\

  $\Omega$ [$^{\circ}$] & 122.43 & 1.12 &-&- \\

  $\omega$ [$^{\circ}$] & 40.065 & 1.118 &-&- \\

  $t_{peri}$ [yr] & 2022.6843 & 8$\times 10^{-4}$\\

    \hline \hline

   \multicolumn{5}{|c|}{\rule{0pt}{2ex}\textbf{S55}}\\
   \hline

  $a$ [as] & 0.10424 & 5$\times 10^{-5}$ &-&- \\

  $e$  & 0.72980 & 1.8$\times 10^{-4}$ &-&- \\

  $i$ [$^{\circ}$] & 159.59 & 0.17 &-&- \\

  $\Omega$ [$^{\circ}$] & 319.43 & 0.97 &-&- \\

  $\omega$ [$^{\circ}$] & 327.77 & 0.93 &-&- \\

  $t_{peri}$ [yr] & 2009.4738 & 7.6$\times 10^{-3}$ &-&- \\

  \hline

\end{tabular}

  \caption{Best-fit parameters of the four-star fit determining $f_{SP}$ (Section \ref{sec:results_schw}). The orbital elements are meant in the sense of osculating orbit parameters, using a conversion time close to the respective apocenter times, i.e., 2010.35 for S2, 1977 for S29, 2000 for S38, and 2012 for S55. The reference frame parameters $x_0$ and $y_0$ refer to the epoch 2000.}
  \label{table_app}
\end{table}

\begin{figure}
   \centering
    \includegraphics[width=0.5\textwidth]{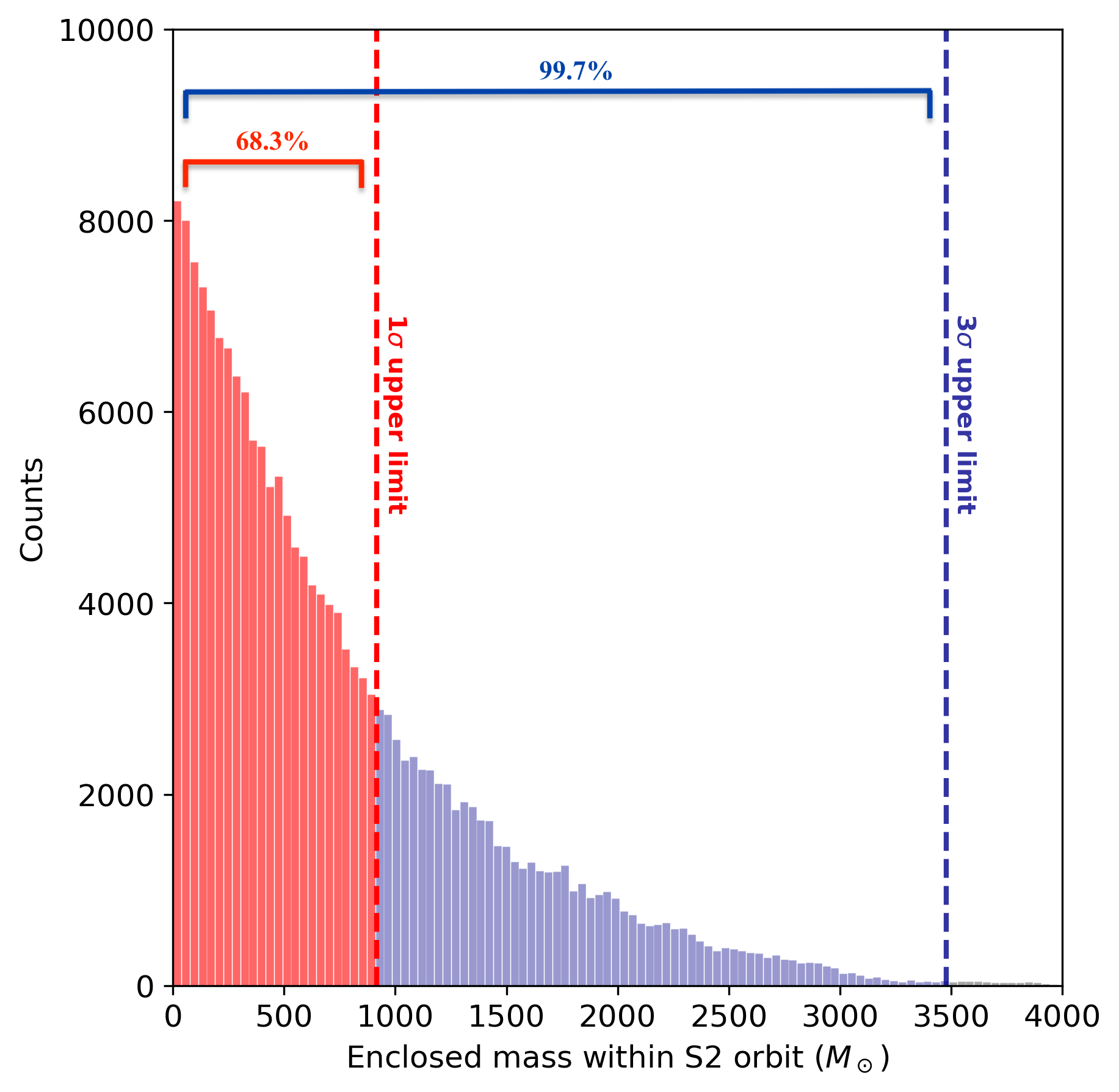}
   \caption{Example of posterior distribution on the enclosed mass within S2's orbit, for an MCMC analysis with $200 \, 000$ realizations, showing how the $1 \sigma$ and $3 \sigma$ upper limits are derived. This example corresponds to the case of a multi-star fit with the stars S2, S29, S38, S55 for a power-law density profile with slope $s=-2.2$.
   }
              \label{fig_hist}
    \end{figure}

\begin{figure}
   \centering
    \includegraphics[width=0.5\textwidth]{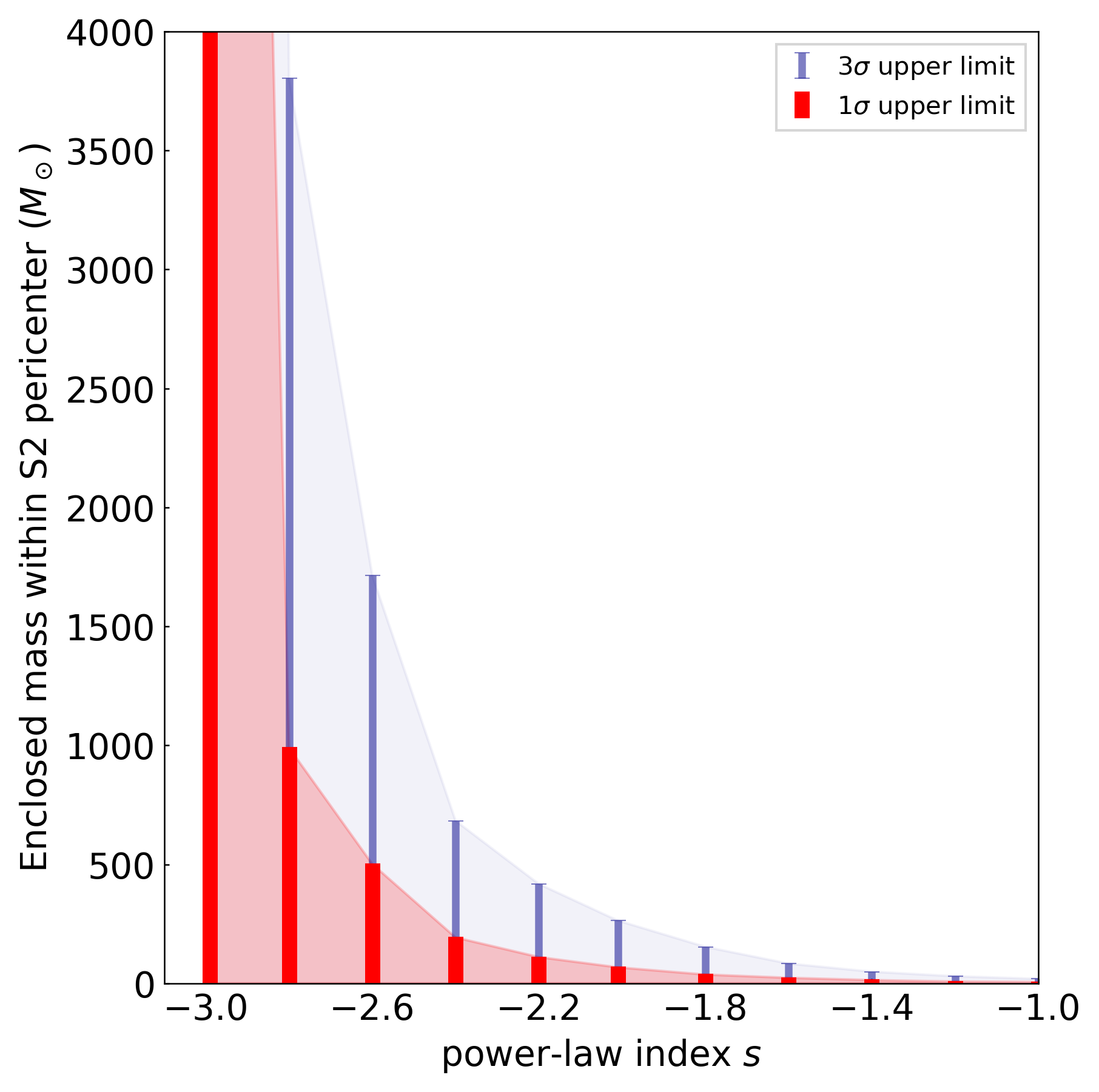}
   \caption{Enclosed mass within S2 pericenter for an extended mass distribution following a power-law density profile with varying slope. In red is plotted the $1 \sigma$ upper limit and in blue the $3 \sigma$ upper limit on this parameter, derived from a multi-star fit with the stars S2, S29, S38, S55.
   }
              \label{fig_mperi}
    \end{figure}

\begin{figure*}
        \centering
        \captionsetup[subfigure]{position=top}
        \begin{subfigure}[b]{0.48\textwidth}
            \centering
            \caption[]
            {\textbf{ Power-law profile}}
            \includegraphics[width=\textwidth]{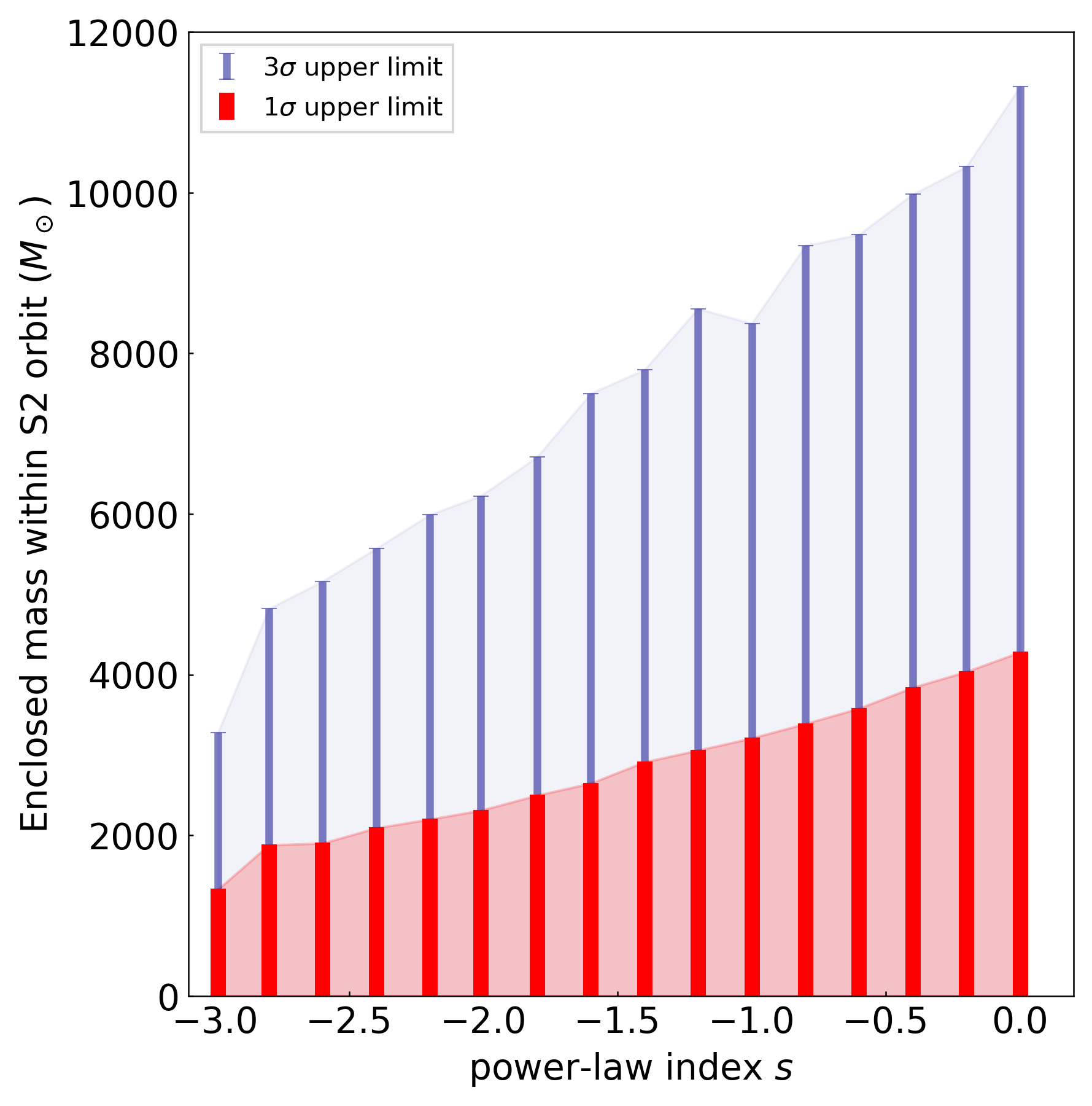}
        \end{subfigure}
        \hfill
        \begin{subfigure}[b]{0.48\textwidth}  
            \centering 
            \caption[]
            {\textbf{Plummer profile}}
            \includegraphics[width=\textwidth]{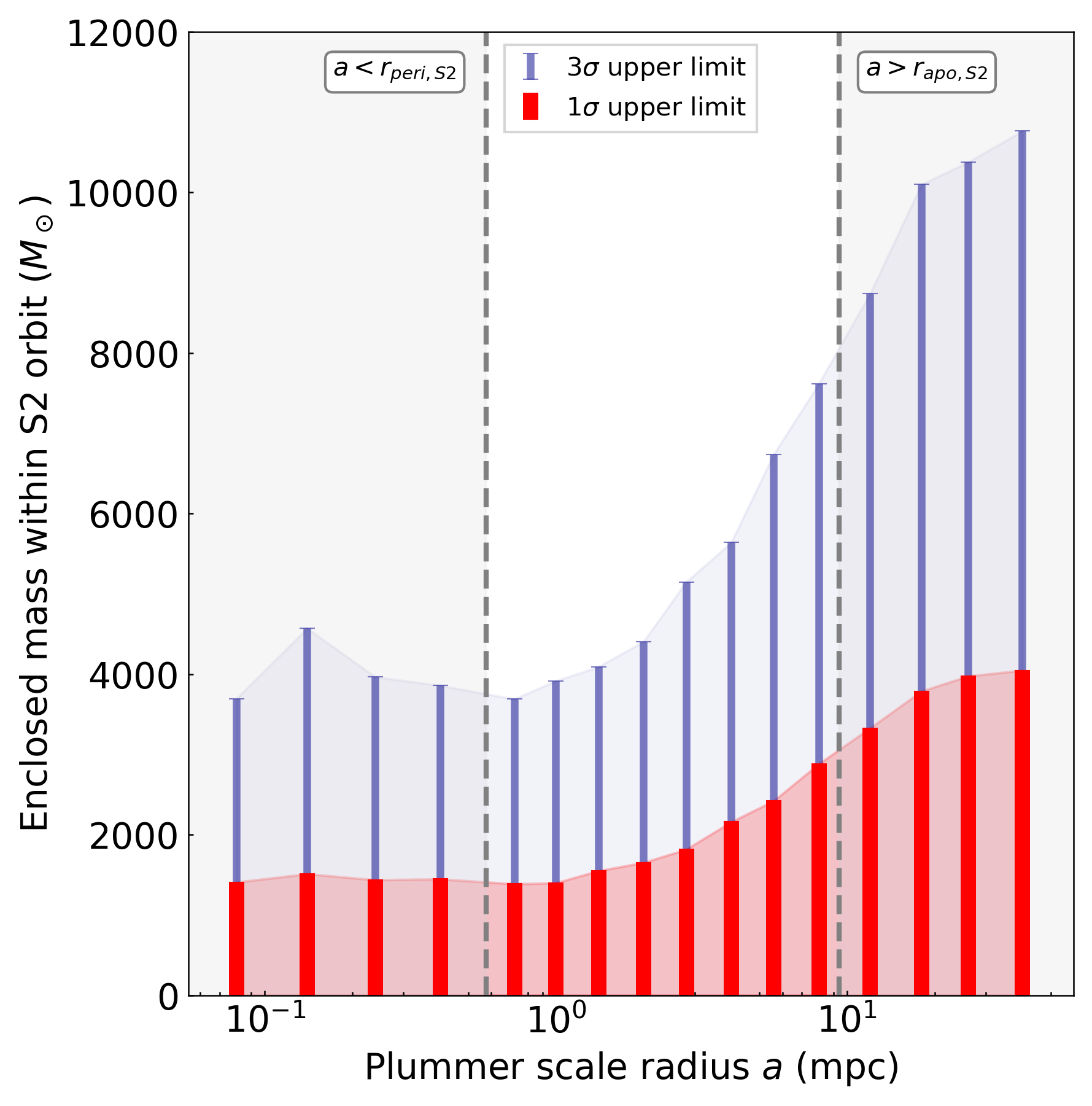}
        \end{subfigure}
        \caption[]
        {Enclosed mass within S2's orbit for an extended mass distribution following a power-law density profile with varying slope (a) and a Plummer density profile with varying scale radius (b). In red is plotted the $1 \sigma$ upper limit and in blue the $3 \sigma$ upper limit on this parameter, derived fitting the orbit of S2.} 
        \label{fig:extmass_s2}
    \end{figure*}

In Table \ref{table_app} we give the best-fit parameters of the four-star fit determining $f_{SP}$ in Section \ref{sec:results_schw}.\\
In Figure \ref{fig_hist} we show an example of posterior distribution on the enclosed mass within S2's orbit, as derived from an MCMC analysis in the case of a multi-star fit with the stars S2, S29, S38 and S55. The $1 \sigma$ upper limit on the enclosed mass corresponds to a 68.3 \% confidence level, the $3 \sigma$ upper limit to a 99.7 \% confidence level.\\
In Figure \ref{fig_mperi} we highlight that the mass enclosed within S2 pericenter is not well constrained with our data, as it is degenerate with the SMBH mass in the fitting. It depends severely on the steepness of the density profile. \\
In Figure \ref{fig:extmass_s2} we show the constraints on the enclosed mass within S2's orbit, obtained fitting the orbit of S2 only. 
Comparing Figure \ref{fig:extmass_4s} with Figure \ref{fig:extmass_s2} it is immediately noticeable that a multi-star fit, namely fitting the data of S2 together with S29, S38 and S55, helps to obtain a much tighter constraint on the enclosed mass within S2's orbit than fitting the data of S2 only.
The constraint is stronger for more compact distributions, namely for a steep power-law distribution and a Plummer profile with a small scale radius, and it becomes weaker for shallower distributions. This can be understood by translating the result on the enclosed mass into the amount of retrograde precession caused by this mass on the stellar orbits: the shallower the distribution, the larger is the amount of extended mass that can lie within the orbit of S2 inducing the same amount of retrograde precession.

\end{appendix}

\end{document}